\documentclass[%
superscriptaddress,
preprintnumbers,
nofootinbib,
amsmath,amssymb,
aps,
twocolumn,
]{revtex4-2}

\usepackage{graphicx}
\usepackage{dcolumn}
\usepackage{bm}
\usepackage{hyperref}
\usepackage{xcolor}
\hypersetup{
    colorlinks,
    linkcolor={blue},
    citecolor={blue},
    urlcolor={blue}
}
\usepackage[normalem]{ulem}

\usepackage{enumitem,amssymb}
\newlist{todolist}{itemize}{2}
\setlist[todolist]{label=$\square$}
\usepackage{pifont}
\newcommand{\diffl}[1]{~{\rm{d}} {#1}}

\newcommand{\thickness}{b}

\usepackage[compact]{titlesec}

\begin{document}

\title{Acoustic radiation from a superconducting qubit:\\ From spontaneous emission to Rabi oscillations
}

\author{Vijay Jain}
\email{vijay.jain@yale.edu}
\affiliation{Department of Applied Physics, Yale University, New Haven, CT 06511}
\affiliation{Yale Quantum Institute, Yale University, New Haven, CT 06511}
\author{Vladislav D. Kurilovich}
\affiliation{Yale Quantum Institute, Yale University, New Haven, CT 06511}
\affiliation{Department of Physics, Yale University, New Haven, CT 06511}
\author{Yanni D. Dahmani}
\affiliation{Department of Applied Physics, Yale University, New Haven, CT 06511}
\affiliation{Yale Quantum Institute, Yale University, New Haven, CT 06511}
\author{Chan U Lei}
\affiliation{Department of Applied Physics, Yale University, New Haven, CT 06511}
\affiliation{Yale Quantum Institute, Yale University, New Haven, CT 06511}
\author{David Mason}
\affiliation{Department of Applied Physics, Yale University, New Haven, CT 06511}
\affiliation{Yale Quantum Institute, Yale University, New Haven, CT 06511}
\author{Taekwan Yoon}
\affiliation{Yale Quantum Institute, Yale University, New Haven, CT 06511}
\affiliation{Department of Physics, Yale University, New Haven, CT 06511}
\author{Peter T. Rakich}
\email{peter.rakich@yale.edu}
\affiliation{Department of Applied Physics, Yale University, New Haven, CT 06511}
\affiliation{Yale Quantum Institute, Yale University, New Haven, CT 06511}
\author{Leonid I. Glazman}
\email{leonid.glazman@yale.edu}
\affiliation{Department of Applied Physics, Yale University, New Haven, CT 06511}
\affiliation{Yale Quantum Institute, Yale University, New Haven, CT 06511}
\affiliation{Department of Physics, Yale University, New Haven, CT 06511}
\author{Robert J. Schoelkopf}
\email{robert.schoelkopf@yale.edu}
\affiliation{Department of Applied Physics, Yale University, New Haven, CT 06511}
\affiliation{Yale Quantum Institute, Yale University, New Haven, CT 06511}

\date{\today}
             
\begin{abstract}
    Acoustic spontaneous emission into bulk dielectrics can be a strong source of decoherence in quantum devices, especially when a qubit is in the presence of piezoelectric materials. We study the dynamics of a qubit coupled to an acoustic resonator by a piezoelectric film. By varying the surface topography of the resonator from rough to polished to shaped, we explore the crossover from fast decay of an excited qubit to quantum-coherent coupling between the qubit and an isolated phonon mode. Our experimental approach may be used for precision measurements of crystalline vibrations, the design of quantum memories, and the study of electro-mechanical contributions to dielectric loss.
\end{abstract}

\maketitle

 \section{Introduction}
Circuit quantum electrodynamics (cQED) is a versatile platform for universal quantum computation~\cite{wallraff04,schoelkopf08,blais21} and the design of hybrid quantum architectures~\cite{clerk20}. Fast, multi-cavity control is enabled by a transmon qubit's large electric dipole moment, which can strongly couple to several microwave modes simultaneously~\cite{gao18}. In the emerging domain of quantum acoustics, a superconducting qubit can efficiently couple to collective vibrations -- phonons -- to prepare non-classical states of sound and coherently exchange quantum excitations~\cite{oconnell10,gustafsson14,chu17,chu18,satzinger18,arrangoiz19, bild2022}. Given that the speed of sound is much slower than that of light, phonons in crystalline media may form a high-density quantum random access memory in a compact form factor with a transmon serving as a non-linear mixing element to interface multiple acoustic modes~\cite{hann19}. However, coupling a qubit to an acoustic medium with many degrees of freedom can lead to rapid decay if the modes are either very lossy or  {if coupling to a continuum of modes produces unintended acoustic radiation.}

The spontaneous emission rate of a quantum emitter is determined by the density of states (DOS) in the environment. According to Fermi's Golden Rule, the decay rate $\gamma$ of a two-level system from spontaneous emission reflects the DOS $D(\omega)$ at its transition frequency $\omega$ and its coupling  {rate} $g$ to those states, or $\gamma=2\pi |g|^{2} D(\omega)$. A transmon qubit in free space would have a sub-micro second lifetime because its large size makes it an efficient radiator to the electro-magnetic continuum. However, embedding the qubit in a high Q{-factor} microwave cavity suppresses the continuum DOS by several orders of magnitude when the qubit is strongly detuned from the cavity resonance~\cite{houck08}. This leads to an inverse Purcell effect~\cite{kleppner81}, where the qubit is protected from radiative decay such that other non-radiative mechanisms begin to dominate.

To realize the potential of quantum acoustic platforms, it is necessary to suppress unintentional acoustic radiation from the qubit for practical integration in quantum computers. In  quantum acoustics, qubits are coherently coupled to surface and bulk acoustic waves and to phononic defect cavities using piezoelectric materials~\cite{luepke22,wollack21,andersson22,kervinen20}. While strong coupling has enabled single-phonon control, the lifetimes of qubits in current piezo devices can be up to two orders of magnitude lower than those of conventional transmons. Since the qubit footprint is much larger than the acoustic wavelength, it is reasonable to assume that a piezoelectric transducer may unintentionally radiate into the acoustic continuum because of the large number of modes in the substrate at GHz frequencies~\cite{scigliuzzo20}. To improve qubit coherences, we must build upon linear circuit models~\cite{arrangoiz16}
 {and develop a microscopic treatment of phonon-qubit coupling to formulate a complete picture of acoustic radiation.}

In this work, we investigate acoustic radiation of a transmon with piezoelectric transducers on a high overtone bulk acoustic wave resonator (HBAR) made of thin film aluminum nitride on sapphire. By modifying the surface topography, we systematically vary the acoustic density of states (ADOS) and explore three unique regimes of acoustic radiation. 
First, we couple the qubit to a resonator with a roughened backside and observe that the qubit's lifetime shortens by two orders of magnitude. 
Disorder on the scale of the acoustic wavelength makes the qubit irreversibly emit its energy into a continuum of phonon modes.
Next, by coupling the qubit to a flat acoustic resonator, we observe a modulation of its lifetime which is consistent with discrete bands in the expected ADOS.
Finally, by shaping the piezoelectric transducer, we form acoustic bound states whose spectral signatures we observe in the qubit's decay and to which we can coherently couple. Our results indicate the importance of acoustic spontaneous emission in designing hybrid quantum systems and develops an experimental approach for controlling this source of decoherence.

\begin{figure}
    \centering
    \includegraphics[width=0.49\textwidth]{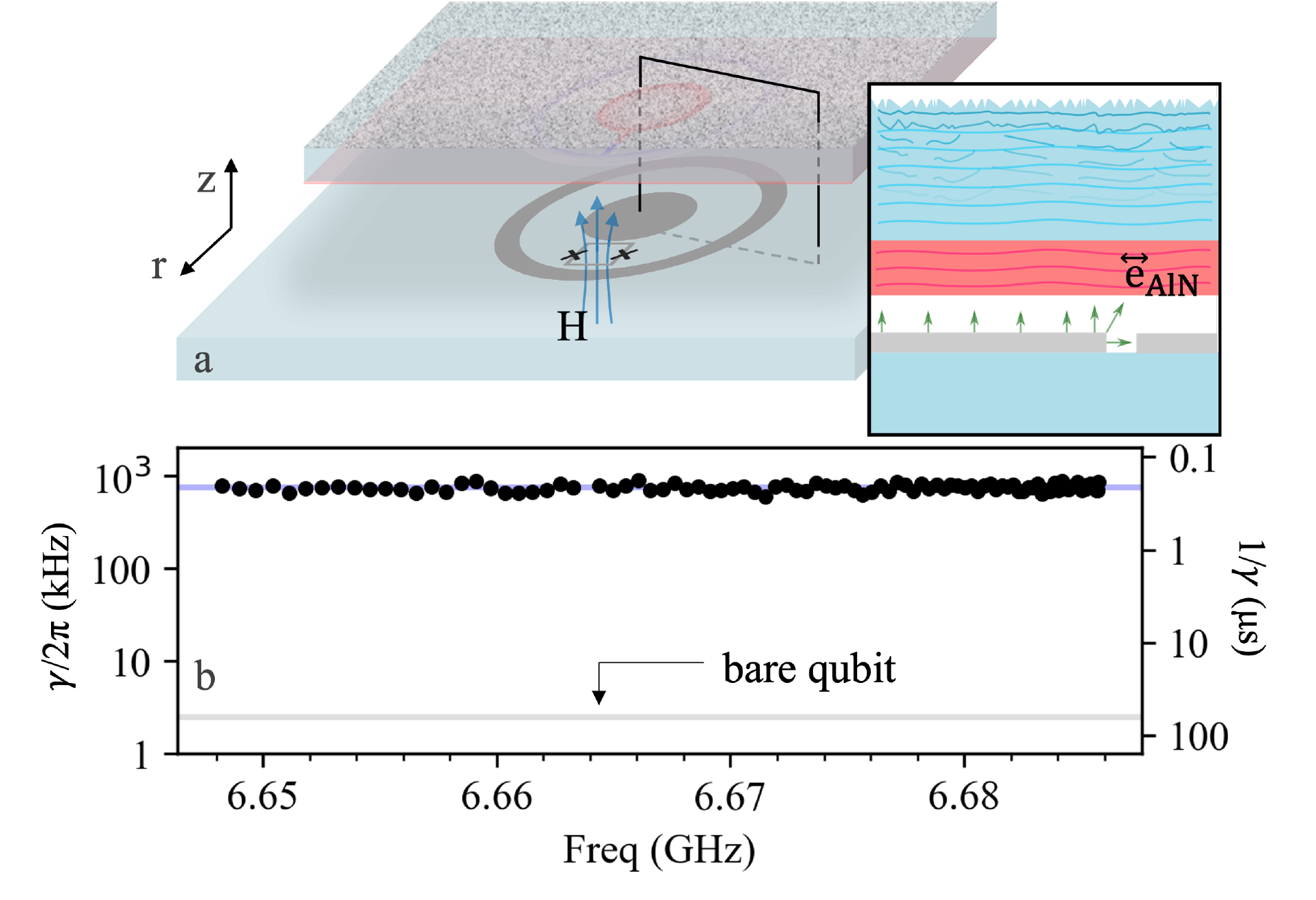}
    \caption{{\bf Acoustic spontaneous emission from a transmon.} (a) An illustration of the sample geometry. Stacked above the center of the concentric transmon is a chip with $1~\mu$m of aluminum nitride on $430~\mu$m of sapphire, which has a roughened backside. The chip entirely covers the qubit. An external magnetic field $H$ is applied (field lines in blue) to flux tune the qubit. (Inset) The electric field of the transmon mode excites an acoustic wave in the resonator, via the piezoelectric effect in AlN. The wave scatters diffusely at the roughened backside. (b) Measurements of the qubit's decay $\gamma$ as a function of its transition frequency. A blue line indicates the analytical estimate using Eq.~\eqref{eq:free_space} when $e_{33}=0.52~\rm{C/m}^{2}$ and a dashed gray line the decay rate of a control qubit $\gamma_{\rm ctrl}$.
}
    \label{fig:spontaneous-emission}
\end{figure}

 \section{Measuring Acoustic  Radiation}
To study acoustic  {radiation}, we {track the dynamics} of a flux-tunable qubit as a function of frequency when coupled to different acoustic environments. Our experimental platform consists of a flip-chip assembly, in which one chip contains  {the qubit and its metallic traces} and the other the HBAR. The qubit has a concentric capacitor layout that shunts a  {pair} of parallel Josephson junctions, resulting in a transition frequency at approximately $\omega_{0}/2\pi = 5-6$~GHz.
Flux threading the square loop is used to tune the qubit's frequency over a broad range; strong, off-resonant voltage drives are used to quickly Stark shift the qubit's frequency in a narrow range. An adjacent meandering stripline is hanger coupled to a transmission line for dispersive readout. The HBAR is formed from sapphire with a film of aluminum nitride $\thickness_{p}=1~\mu$m thick;
aluminum spacers are deposited to form a $1~\mu$m vacuum gap between the qubit and the piezoelectric film. The assembly is mounted in a superconducting package at the base of a dilution refrigerator.

We conduct experiments using three different styles of HBARs which systematically vary the ADOS. First, we use a single-side polished sapphire chip with thickness $\thickness = 430\,{\rm \mu m}$, whose backside has a roughness of approximately $1~\mu$m-rms (shown in Fig.~\ref{fig:spontaneous-emission}a). Next, we repeat the same with a $\thickness = 100\,{\rm \mu m}$ double-side polished sapphire chip. Finally, we make two separate samples, in which a $\thickness = 100\,{\rm \mu m}$ thick sapphire chip is patterned with an aluminum nitride cylinder of diameter $250~\mu$m or a dome formed through a vapor-phase reflow process~\cite{kharel2018} and dry-etching of  the residual piezoelectric.

 \subsection{Radiation into acoustic free space\label{sec:free_space}}

To understand the impact of acoustic radiation on qubit coherence, we begin by investigating the coupling of the qubit to a quasi-continuum of phonon modes. Here, we employ an HBAR with a roughened backside, with roughness on the order of the acoustic wavelength. {A qubit is first brought into the excited state with a fast pulse. Field oscillations within the excited qubit couple (via the piezoelectric effect) to an acoustic wave that traverses the bulk and diffusely scatters from roughened back surface.
Hence, this acoustic excitation does not coherently return to the transducer, leading to spontaneous emission analogous to that of a qubit coupled to acoustic free space.}
Indeed,  {measurements reveal} a fast decay time of $T_{1}=0.21~\mu$s that does not depend on the qubit's frequency in a $40$~MHz bandwidth (a resonator with the same nominal thickness would have a 13 MHz free spectral range). By comparison, a control qubit, which lacks the top HBAR chip entirely, has a lifetime $T_{1}=62~\mu$s. Therefore, acoustic spontaneous emission is the dominant source of decoherence, occurring at a rate $\gamma/2\pi = 750$~kHz.

We model the roughened HBAR as an acoustic half-space to analytically estimate the radiation rate.
The main contribution to the spontaneous emission rate in our device comes from the emission of longitudinal waves.
The coupling between the qubit and a longitudinal phonon $\hbar g = - \int d^3{\bm r}\,\sigma_{zz}({\bm r}) s_{zz}({\bm r})$ is determined by the overlap of stress $\sigma_{zz}({\bm r})= e_{33} E_{z}({\bm r})$ generated by the qubit's electric field $E$ in the piezoelectric film, and strain $s_{zz}({\bm r}) = s_{0} e^{i{\bm k}_\perp \cdot {\bm r}_\perp} {\rm sin} (k_z z)$ associated with the phonon. Here, $e_{33}$ is the piezoelectric modulus of AlN and $s_0$ is the zero-point strain amplitude. Applying Fermi's Golden Rule, we find~\cite{supplinfo}:
    \begin{equation}\label{eq:free_space}
        \gamma_{\rm fs}(\omega_0)=\frac{2\pi}{\hbar}\frac{4e_{33}^{2}}{\pi v_{l}\rho\omega_0}{\rm sin}^4\Bigl(\frac{\omega_0 \thickness_{p}}{2v_{l}}\Bigr)\int d^{2}\bm{r}_{\perp}E_{z}^{2}(\bm{r}_{\perp}),
    \end{equation}
where $\rho = 4\cdot 10^3\,{\rm kg/ m^3}$~is the density of sapphire and $v_{l}=11.1$~km/sec is the longitudinal wave velocity.
In the derivation, we assumed that the electric field does not change appreciably across the thickness of the piezoelectric film. To find $E_z({\bm r}_\perp)$, we use an HFSS simulation, which results in $\int d^{2}\bm{r}_{\perp}E_{z}^{2}(\bm{r}_{\perp}) = 5.2\cdot 10^{-10}\,{\rm V^2}$. For a thin film with $\thickness_{p} \ll \lambda_{\rm ac}$, the decay rate $\gamma_{\rm fs} \propto (\omega_0/v_l)^{3}$, analogous to radiation by a dipole in free space.
In our experiment, however, the piezoelectric film thickness $\thickness_p$ is close to half the acoustic wavelength ($\lambda_{\rm ac}/2= 0.9~\mu$m) at the qubit's transition frequency, {yielding} $\textrm{sin}(\omega_{0}\thickness_{p}/(2v_{l}))\approx 1$.

\subsection{Flat acoustic resonator}
\begin{figure*}[t]
    \centering
    \includegraphics[width=0.94\textwidth]{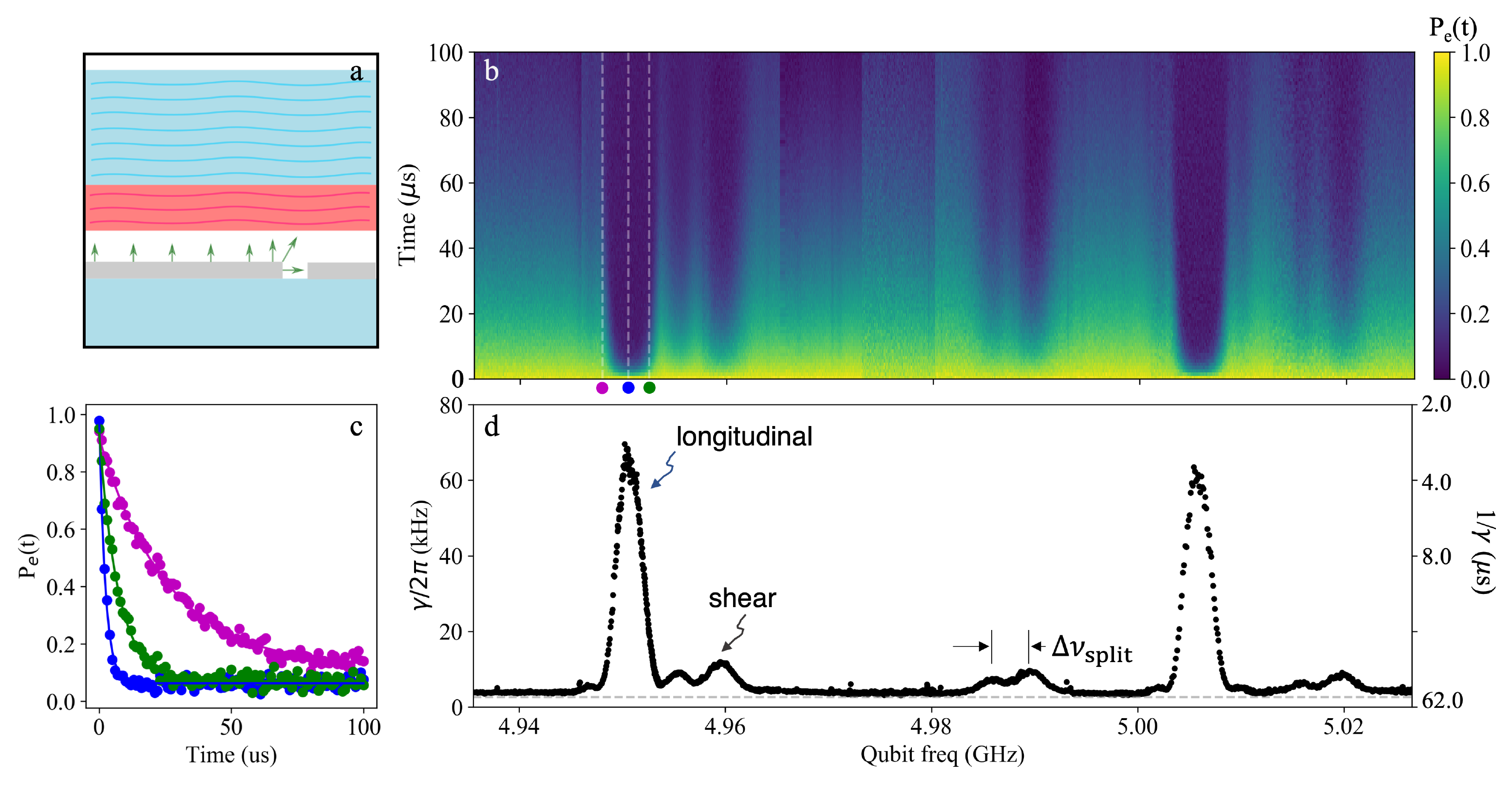}
    \caption{{\bf Dynamics of the qubit coupled to a flat acoustic resonator.} 
    (a) As opposed to the inset of Fig.~\ref{fig:spontaneous-emission}, the backside of the acoustic resonator is flat.
    (b) Time-dependence of the excited state population of the qubit $P_e(t)$ at different qubit frequencies $\omega_0$ ($P_e(t)$ is shown with the color). 
    (c) Line-cuts of the plot in panel (b) at three different qubit frequencies. Data-points in blue, green, and magenta correspond to the frequencies indicated in panel (b) with circles of the respective colors. The time-dependence of $P_e(t)$ is well-described by a single decaying exponent (solid lines).
    (d) Measured decay rate $\gamma(\omega_0)$ of the qubit at each frequency. The grey dashed line is the decay rate of the control sample $\gamma_{\rm ctrl}$. The peaks in $\gamma(\omega_0)$ correspond to the resonances between the qubit and the standing acoustic waves.
    }
    \label{fig:exp-data}
\end{figure*}

The estimate for $\gamma_{\rm fs}$ depends sensitively on the value of piezoelectric modulus $e_{33}$, which we have not measured independently.
Upon using in Eq.~\eqref{eq:free_space} the value $e_{33} = 1.4\,{\rm C / m^2}$ previously reported for polycrystalline AlN thin films \cite{muralt1999, muralt2001}, we find $\gamma_{\rm fs} = 2\pi \cdot 5.9\,{\rm MHz}$.
While this estimate agrees with the measured rate within an order of magnitude, it nonetheless overestimates the experimentally observed rate by a factor of $\approx 7$.
We can bring the theory and the data into agreement if we use $e_{33} = 0.52\,{\rm C / m^2}$. 
Although this value is lower than that for thin AlN films at room temperatures (by a factor of $\sim 2.5$), it is close to the $e_{33}$ extracted from measurements of the coherent coupling to a discrete phonon mode in other quantum acoustic devices \cite{chu17}.

The large difference between the lifetimes of a control qubit ($T_1 = 62$~$\mu$s) and a qubit in the roughened HBAR  {structure} ($T_1 = 0.20$~$\mu$s) verifies the impact of acoustic spontaneous emission on coherence.  {This disparity} could be even more dramatic in other quantum acoustic designs. From the participation ratio $p_{r} = \epsilon_{0}\epsilon \int \diffl{V} E_z^{2} / \hbar \omega$, we estimate that the qubit stores $1\%$ of its electrical energy in the piezoelectric. Alternate designs with stronger in-film electric-field strengths or piezoelectric coupling $e_{33}$ may  {cause} the qubit's lifetime to drop to nanosecond timescales.

Next, we modify the qubit's acoustic environment with a flat acoustic resonator to investigate the effect of a reduced continuum ADOS. Specifically, we allow the formation of longitudinal resonances, which creates sharp bands of ASE on the background of otherwise suppressed (by two orders of magnitude) emission. As before, the qubit's $E$-field couples to a continuous piezoelectric film on the bottom face, but now  {the} resonator has a smoothly polished top face. In this case, the qubit's decay rate spectrum becomes frequency dependent with a series of features, see Fig.~\ref{fig:exp-data}. A sharp decay peak of $\gamma_q / 2\pi = 70$~kHz repeats with a  {periodicity} of $v_l / (2\thickness) = 55.5$~MHz. This free spectral range corresponds to round-trip travel time of the longitudinal wave across the acoustic resonator.
A weaker decay split-peak  {in the decay rate} repeats with $v_{\rm sh} / (2\thickness) = 31$~MHz, corresponding to the round-trip time of a shear wave ($v_{\rm sh}=6.1$~km/sec is the shear wave velocity).
Away from the peaks, the decay rate $\gamma / 2\pi = 3.4~$kHz is close to that of the control qubit, $\gamma_{\rm ctrl} / 2\pi = 2.6~$kHz.

To analyze the observations, we use the same model for the piezoelectric interaction of the qubit with phonons as in Sec.~\ref{sec:free_space}. The only modification is that now the spectrum of $k_z$ wavevectors is discrete, $k_z=n\pi/b$.
{The HBAR chip acts as a thick multimode} waveguide; the wave spectrum in it is characterized by the in-plane wavevector $k_\perp$ and the standing wave overtone number $n$ for each of the wave polarizations.
The largest peaks in Fig.~\ref{fig:exp-data}c correspond to the longitudinal mode with the spectrum $\omega_{n}^{2}(k_\perp)=v_{l}^{2} (n\pi/\thickness)^{2} + v_{\perp}^{2} k_{\perp}^{2}$.
The subset of wavevectors $k_\perp$ of phonons efficiently interacting with the qubit are limited by diffraction to $k_\perp\lesssim \pi/a$; here $a$ is the qubit size, $\pi/a\sim (100\,{\rm \mu m})^{-1}$.
The respective mode frequencies $\omega_n(k_\perp)$ form narrow intervals of width $\omega_{\rm diff} \sim v_{\perp}^2  \pi^2/ (\omega_n a^2) \sim 2\pi \cdot 10$~kHz adjacent to the standing wave frequencies $\omega_n = \pi n v_l / \thickness$. 
The smallness of $\omega_{\rm diff}$ in comparison to the frequency separation between standing waves  {($\pi v_l / b = 55$~MHz)} stems from the immense difference between the size of the qubit and the acoustic wavelength. 

Application of Fermi's Golden Rule to the phonon emission would lead one to conclude that a substantial contribution of phonons to the qubit decay rate occurs only within the said narrow frequency intervals. However, the applicability condition for Fermi's Golden Rule is $g\ll\omega_{\rm diff}$, i.e., the phonon-qubit coupling must be sufficiently weak. Here $g$ is the strength of coupling between the qubit and a standing wave.
{Using the value of the piezo-modulus $e_{33}$ found in Sec.~\ref{sec:free_space}, we estimate $g \sim 1$~MHz.}
The coupling exceeds the diffraction linewidth $\omega_{\rm diff} \sim 10$~kHz by two orders of magnitude. This renders Fermi's Golden Rule inapplicable and calls for a comprehensive solution of the qubit-phonon dynamics problem \cite{supplinfo}. We find that the qubit, which is tuned to a resonance with a standing wave, should undergo vacuum Rabi oscillations with a frequency $\Omega_{\rm R} = 2g$.
Rabi oscillations with $\sim 1$ amplitude {would} occur for the qubit frequencies spanning the range $|\omega_0 - \omega_n| \lesssim g$. The phonon continuum $\omega_{n}^{2}(k_\perp)=v_{l}^{2} (n\pi/\thickness)^{2} + v_{\perp}^{2} k_{\perp}^{2}$ leads to the oscillations decay. We find the decay rate $\gamma_{\rm R} \sim \sqrt{\omega_{\rm diff} g} \sim 100$~kHz at the resonance $\omega_0=\omega_n$. The closeness of the qubit frequency to the phonon mode threshold makes the character of the decay peculiar: the population of the qubit excited state drops down to $1/4$ rather than $0$. Further relaxation occurs via other, slower mechanisms. The estimated $\gamma_{\rm R}$ agrees by the order of magnitude with the observed peak decay rate, while the peak width agrees with estimated value of $g$ reasonably well. We must emphasize, however, that no Rabi oscillations were observed in the studied device, see Fig.~\ref{fig:exp-data}. The decay is well-described by a single exponent, in contrast to the theoretical prediction.

{In absence of observable Rabi oscillations,} it is tempting to explain the experimental findings using Fermi's Golden Rule. The explanation requires two assumptions. First, to account for the measured magnitude of $\gamma$, the coupling strength must be $g \sim 100$~kHz, an order of magnitude smaller than our theoretical estimate, $g\sim 1$~MHz. Second, an explanation of the spectral broadening $\kappa \sim 1~$MHz requires one to assume the device inhomogeneity  {such as a residual surface roughness or the crystalline defects in the bulk.} Currently, we do not see a justification for these assumptions.

To conclude this section, we note another interesting feature in the data of Fig.~\ref{fig:exp-data}: the splitting of the shear wave resonances into two peaks corresponding to the two shear polarizations.
The splitting likely originates from a slight misalignment between the $z$-axis of the device (along which the waves are launched) and the $c$-axis of sapphire crystal.
Were the two axes perfectly aligned, the two shear waves would have had the same propagation velocities in the $z$-direction, and the identical sets of the standing wave frequencies.
The misalignment {of these axes} by $\Delta \theta$ results in a relative difference of the shear wave velocities and corresponding standing wave frequencies, $\Delta \omega_{\rm split} / \omega_0 = \Delta v_{\rm sh} / v_{\rm sh} \propto \Delta \theta$. We estimate \cite{supplinfo} that the observed splitting can be explained by  {a misalignment angle} $\Delta \theta = 0.15^\circ$; this estimate is close to the alignment error $\pm 0.1^{\circ}$ in the specifications of the sample.

 \subsection{Topographic deformation}
\begin{figure}[t!]
    \centering
    \includegraphics[width=0.49\textwidth]{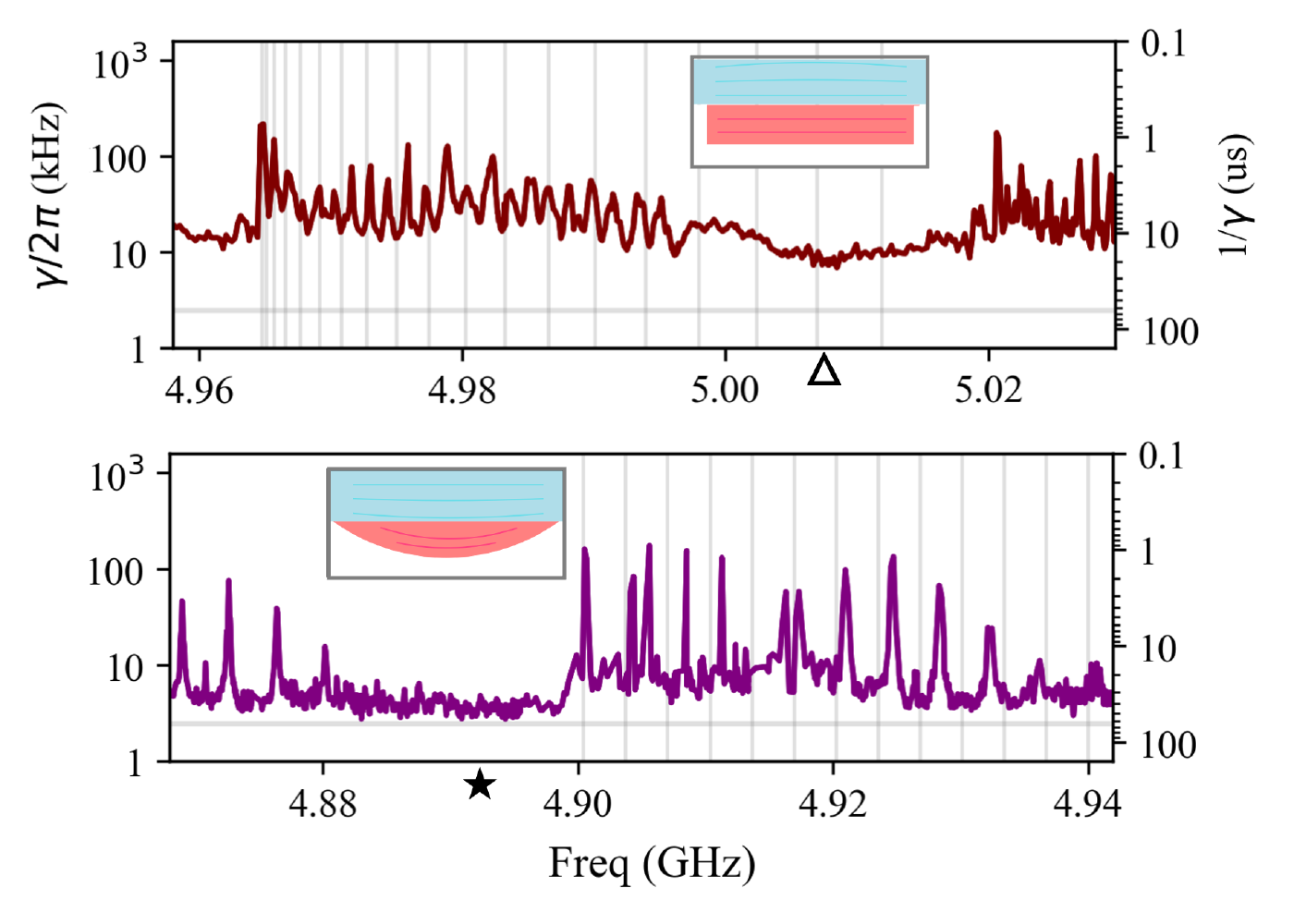}
    \caption{{\bf Topographic modification.} The qubit's decay $\gamma=1/T_{1}$ is measured using fast Stark shifts in two different experimental geometries. The aluminum nitride thin film is etched away to reveal either a cylinder (top) or dome (bottom) transducer approximately the width of the center pad. Here, several discrete modes are visible as the peaks in the decay rate; some of these modes are strongly-coupled to the qubit. Grey vertical lines indicate the calculated mode frequencies.
    }
    \label{fig:topographic-modification}
\end{figure}

{Finally, by changing shape of the acoustic resonator, we create a series of discrete phonon modes that produce a distinctly different spectrum of spontaneous emission.}
We fabricated two devices with different transducer shapes depicted schematically in Fig.~\ref{fig:topographic-modification}.
In the first device, we pattern a cylinder, etch away the remaining piezoelectric film, and stack it above the center conductor of the qubit.
The radius of the cylinder $r = 125\,\mu$m is chosen such that the transducer is slightly smaller than the qubit's center conductor;
this reduces the sensitivity of the device performance  {to the sharply-varying fields at the edges of the conductors.}
In the second device, we pattern a dome-shaped transducer with $r = 125$~$\mu$m and radius of curvature ${\cal R} = 7.8$~mm.
The measured dependence of the qubit's decay rate on its frequency is presented for both devices in Fig.~\ref{fig:topographic-modification}.

A striking feature of Fig.~\ref{fig:topographic-modification} is the presence of tightly-packed peaks in the qubit's decay rate whose spacing $\delta\nu \sim 1$~MHz is much smaller than 
than the fundamental frequency of the bulk standing wave $v_l / (2\thickness) = 55.5$~MHz.
We attribute  {these} peaks to the resonances between the qubit and {the long-lived acoustic ``bound'' states. Such bound states (or leaky modes) are formed in the volume of the chip above the transducer by the lateral confinement.}
To verify the origin of the peaks, we solve the Christoffel wave equation for the two geometries at hand \cite{supplinfo}, and compare the resulting spectra of bound state frequencies with the positions of peaks in $\gamma(\omega_0)$.
For each bulk wave overtone $n$, we find a series of bound states distinguished by the transverse wave number, with frequency spacing $\delta \nu \ll v_l/(2\thickness)$. The smallness of the spacing stems from the respective smallness of the acoustic wavelength $\lambda_{\rm ac}\ll {\cal R}, \thickness, r$.
The computed bound state frequencies are depicted in Fig.~\ref{fig:topographic-modification} by the vertical grey lines.

We can also estimate the number of resolvable bound states for each overtone $n$. Due to the smallness of $\lambda_{\rm ac}$, their frequencies $\omega_n^2(k_\perp)=v_l^2(n\pi/(\thickness+\thickness_p))^2+v_{\rm \perp}^2k_\perp^2$ are accurately estimated by making $k_\perp$ discrete. For example, for the cylindrical transducer
$k_\perp$ changes in steps of width $\sim \pi / r$.
The bound states are resolvable as long as the respective $\omega_n(k_\perp)$ 
are lower than the edge of the corresponding overtone's band, $v_ln\pi/\thickness$. This yields
19 resolved states. A similar calculation for the dome gives
15 resolved states~\cite{supplinfo}. These numbers are in a reasonable agreement with observations. 

The decay rate decreases upon detuning the qubit away from the resonances, reaching a minimum background value of $6.6$ ($\bigtriangleup$) and $2.8$~kHz ($\star$) for the cylinder and dome geometries, respectively.
Subtracting the control qubit's decay rate $\gamma_{\rm ctrl}/2\pi=2.6$~kHz, we find the contribution of acoustic spontaneous emission is $\gamma_{\rm rad}/2\pi = 3.2$~kHz for the cylinder geometry and $\lesssim 0.2$~kHz for the dome. The elevated radiation from the cylinder stems from diffraction of acoustic waves in the bulk on the transducer's sharp edge; the dome and its tapered edges, by contrast, suppresses coupling to propagating waves in the bulk. While dielectric losses in the transducer could also contribute to differences from the control, we note that dielectric participation in these two samples is similar, while their $\gamma_{\rm rad}$ differs by at least an order of magnitude. This verifies that acoustic radiation is more likely than dielectric loss to limit coherence.

 \subsection{Acoustic bound state spectroscopy}
\begin{figure}[t!]
    \centering
    \includegraphics[width=0.48\textwidth]{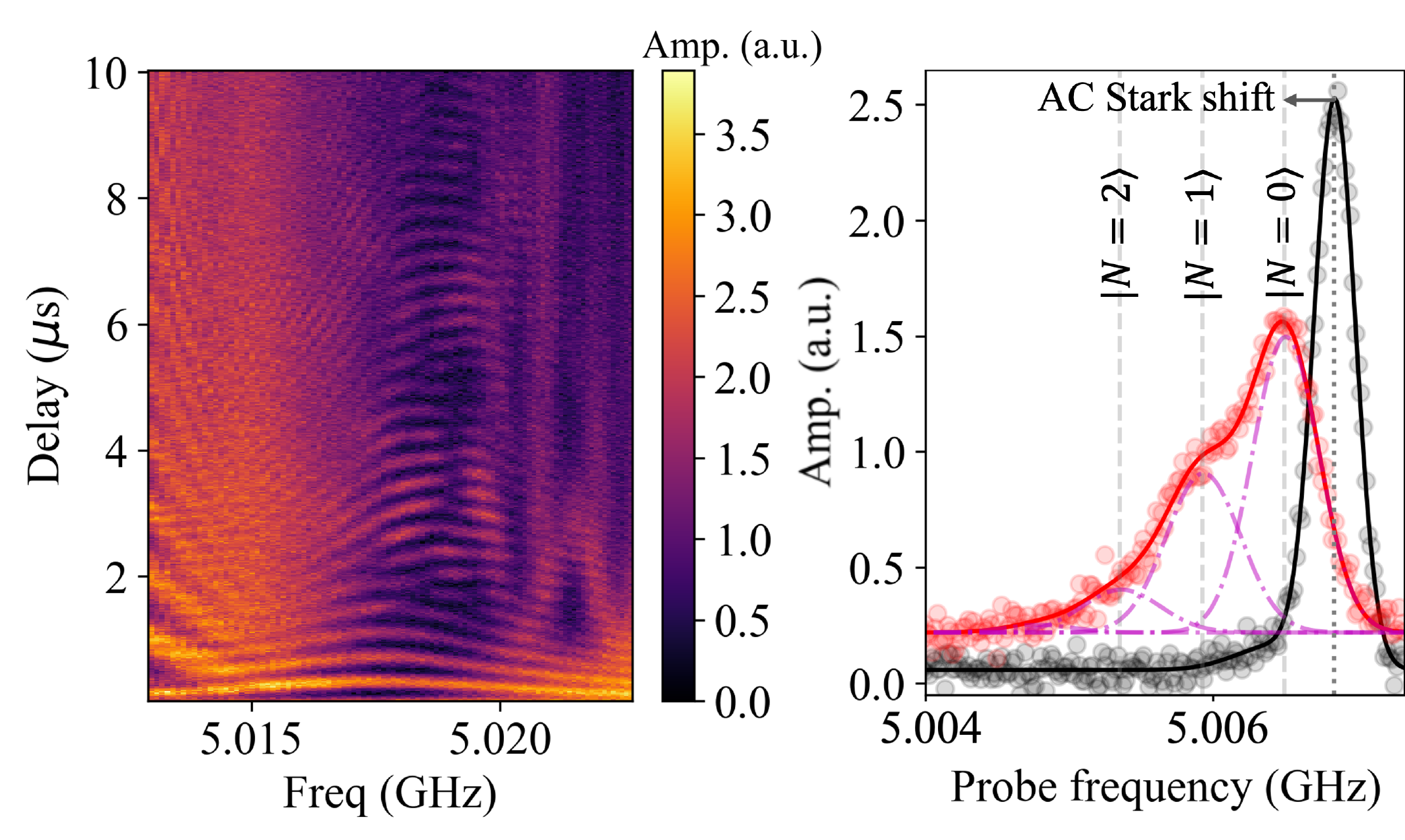}
    \caption{{\bf Spectroscopy of cylindrical HBAR.} (Left) Vacuum Rabi oscillations between the transmon qubit and a phonon. The color reflects the qubit’s excited state probability as a function of time. The oscillations reveal a coupling rate of $g/2\pi = 1.35$~MHz. The respective data for the dome geometry is presented in \cite{supplinfo}. (Right) A measurement of the dispersive shift to the qubit from a driven acoustic mode at $\omega_{\rm ph}/2\pi = 5.0175$~GHz. Here, the qubit is detuned by $\Delta/2\pi = -10.5$~MHz and its transition frequency ($\omega_{0}=\tilde{\omega}_{0} + N \chi$) and linewidth ($\gamma_{q}=2\gamma_{2}+N\kappa$) depend on the number of excitations $n$ in the displaced acoustic mode~\cite{gambetta06}. The linewidth of the $|N=1\rangle$ peak is $\Delta\gamma=21.8$~kHz broader than that of the $|N=0\rangle$ peak, which bounds the acoustic lifetime to $T_{1}^{\rm ph} = 7.3~\mu$s. \label{fig:disp-shift}}
\end{figure}

The discrete nature of the acoustic bound states is confirmed by the presence of multiple vacuum Rabi oscillations in the time domain. Plotted in Fig.~\ref{fig:disp-shift}a is the time evolution of the qubit's excited state population in the cylindrical geometry.
We observe a clear pattern of vacuum Rabi oscillations. 
At $\omega_{0}/2\pi=5.0175$~GHz, the oscillations occur at a vacuum Rabi rate $\Omega_{\rm R}/2\pi = 2.7$~MHz.
By identifying $\Omega_{\rm R} = 2g$, we infer $g / 2\pi = 1.35$~MHz for the coupling of the qubit-phonon coupling. 
This value agrees reasonably well with our estimate, $g / 2\pi = 1\,{\rm MHz}$, for the qubit interacting with the principal transverse mode in the cylinder (in the estimate we use $e_{33} = 0.52\,{\rm C/m^2}$ for the piezoelectric coefficient of AlN, as deduced in Sec.~\ref{sec:free_space}; see \cite{supplinfo} for details).
We note that the observed pattern of oscillations is distorted from its standard chevron shape.
The reason for the distortion is the high value of the coupling strength: $g$ is close to the frequency spacing between the bound states $\delta\nu \sim 1$~MHz, so the qubit hybridizes with multiple acoustic modes simultaneously.

Next, we find the acoustic bound state decay rate $\kappa$ using the dispersive coupling to the qubit.
First, by flux-tuning the qubit, we park its frequency $\omega_0$ sufficiently close to $\omega_{\rm ph}$. To enter the dispersive coupling regime, we choose the detuning $\Delta / 2\pi = (\omega_0 - \omega_{\rm ph})/2\pi = -10.5$~MHz that exceeds $g/2\pi = 1.35$~MHz.
Then, we apply a Gaussian pulse resonant with the bound state which displaces it into a coherent state – a superposition of Fock states with different phonon numbers $N$.
Due to the dispersive coupling between the qubit and the bound state, $H_{\rm disp} = \frac{1}{2}\chi \sigma_z N$, the spectral line in the qubit microwave response becomes a sum of overlapping Gaussians shifted by $\chi$ with respect to each other, as seen in  Fig.~\ref{fig:disp-shift}.
The dispersive shift from a single phonon can be evaluated as $\chi = 2g^2 / \Delta$ yielding $\chi / 2\pi = -0.4$~MHz. 
The widths of Lorentzians depend on the qubit dephasing rate $\gamma_2$, the acoustic bound state decay rate $\kappa$, and on the phonon number, $\gamma(N) = 2\gamma_2 + \kappa N$~\cite{gambetta06}.
The sensitivity of $\gamma(N)$ to phonon number $N$ allows us to extract $\kappa$ by comparing the linewidths corresponding to different $N$.
We fit the shape of the qubit response by a sum of Lorentzians centered at positions $\tilde{\omega}_0 + N\chi$ with widths $\gamma(N)$ ($\tilde{\omega}_0$ differs from $\omega_0$ due to the induced by the pulse AC Stark shift).
The only fitting parameters are $\gamma_2$, $\tilde{\omega}_0$, and $\kappa$.
We obtain $\kappa/2\pi = 21.8$~kHz for the decay rate of the bound state, or $T_{1}^{\rm ph} = 7.3~\mu$s.

Using this value of $\kappa$, we estimate a single-phonon cooperativity $C = 4g^{2}/(\kappa \gamma_2) = 8.4 \times 10^{3}$. Its value, which is limited in the present work by dephasing in the qubit $\gamma_2/2\pi = 39.8$~kHz,  {represents an increase over the previous HBAR devices \cite{chu17, chu18}}, and is the result of an increase in the coupling strength $g$. Further increase of cooperativity may be achieved by using a transducer with rounded edges (to mitigate the diffraction losses), an acoustic cavity with the reduced surface roughness, and a qubit with reduced dephasing.

 \section{Conclusions and outlook}
The speed of sound is $10^4$ times slower than that of light; this leads to a high density of acoustic modes in crystalline media as compared to that of electromagnetic modes in a resonator of a comparable size.
The high density of acoustic modes is both a resource and a challenge for their use in quantum information applications.
On the one hand, we may build quantum acoustic devices which are much more compact than existing cQED ones. 
On the other hand, the higher density of modes may easily lead to a fast decay of a qubit due to the emission of phonons. 
In this work, we demonstrated the crossover between the regime of fast qubit decay and that of large single-phonon cooperativity; we achieved this by purposefully modifying the device geometry. The relevant length scale for the modifications is determined by the wavelength $\lambda_{\rm ac}$ of a phonon at the qubit frequency. A rough (on the scale $\lambda_{\rm ac}$) surface of the acoustic cavity results in strong phonon diffraction, which leads to a structureless acoustic density of states (ADOS) mimicking acoustic free space for phonons. Expectedly, we observe fast decay of the qubit independent of its frequency, as seen Fig.~\ref{fig:spontaneous-emission}. A smooth on the scale $\lambda_{\rm ac}$ surface limits the spontaneous emission to the set of qubit frequencies resonant with the standing waves in the chip, as seen in Fig.~\ref{fig:exp-data}. Furthermore, in a full analogy with plano-convex resonators in laser physics~\cite{siegman86} shaping of the transducer allows one to isolate discrete long-lived acoustic modes strongly coupled to the qubit. The strong coupling is exemplified by the observed Rabi oscillations, as seen in~Fig.~\ref{fig:disp-shift}.

Our design of the experiments and data analysis went beyond the lumped element circuit models of acoustic modes~\cite{arrangoiz16,ask19} and focused on the wave nature of the coupling. This approach may be extended in several directions.
Going forward, we may use it to investigate radiation into Rayleigh waves at the surface. Additionally, we may assess topographic modifications for use in the design of high-density, multi-mode quantum random access memories based on bulk acoustic wave resonators~\cite{hann19}. Lastly, the technique developed in this work may be used for the precision measurements of the bulk phonon properties and the associated dielectric losses in the qubit substrates~\cite{mueller19}.

\begin{acknowledgements}
We thank Luigi Frunzio and Freek Ruesink for valuable discussions and feedback on the manuscript; Yiwen Chu, Hugo Doeleman, and Uwe von L\"{u}pke for insightful discussions regarding dielectric losses in HBARs; and Nikolay Gnezdilov and Valla Fatemi for initial theoretical discussions. Facilities use was supported by YINQE and the Yale SEAS cleanroom. This research was initially supported by the U.S. Department of Energy, Office of Science under award number DE-SC0019406 and completed under support by the U.S. Department of Energy, Office of Science, National Quantum Information Science Research Centers, Co-design Center for Quantum Advantage (C2QA) under contract number DE-SC0012704.  {R.J.S.
is a founder and equity shareholder 
of Quantum Circuits, Inc.}
\end{acknowledgements}

{\textbf{Disclaimer} This report was prepared as an account of work sponsored by an agency of the United States Government. Neither the United States Government nor any agency thereof, nor any of their employees, makes any warranty, express or implied, or assumes any legal liability or responsibility for the accuracy, completeness, or usefulness of any information, apparatus, product, or process disclosed, or represents that its use would not infringe privately owned rights. Reference herein to any specific commercial product, process, or service by trade name, trademark, manufacturer, or otherwise does not necessarily constitute or imply its endorsement, recommendation, or favoring by the United States Government or any agency thereof. The views and opinions of authors expressed herein do not necessarily state or reflect those of the United States Government or any agency thereof.}

\bibliography{references}




\end{document}


\title{Supplemental Information for ``Acoustic radiation from a superconducting qubit: From spontaneous emission to Rabi oscillations''
}

\author{Vijay Jain}
\email{vijay.jain@yale.edu}
\affiliation{Department of Applied Physics, Yale University, New Haven, CT 06511}
\affiliation{Yale Quantum Institute, Yale University, New Haven, CT 06511}
\author{Vladislav D. Kurilovich}
\affiliation{Yale Quantum Institute, Yale University, New Haven, CT 06511}
\affiliation{Department of Physics, Yale University, New Haven, CT 06511}
\author{Yanni D. Dahmani}
\affiliation{Department of Applied Physics, Yale University, New Haven, CT 06511}
\affiliation{Yale Quantum Institute, Yale University, New Haven, CT 06511}
\author{Chan U Lei}
\affiliation{Department of Applied Physics, Yale University, New Haven, CT 06511}
\affiliation{Yale Quantum Institute, Yale University, New Haven, CT 06511}
\author{David Mason}
\affiliation{Department of Applied Physics, Yale University, New Haven, CT 06511}
\affiliation{Yale Quantum Institute, Yale University, New Haven, CT 06511}
\author{Taekwan Yoon}
\affiliation{Yale Quantum Institute, Yale University, New Haven, CT 06511}
\affiliation{Department of Physics, Yale University, New Haven, CT 06511}
\author{Peter T. Rakich}
\email{peter.rakich@yale.edu}
\affiliation{Department of Applied Physics, Yale University, New Haven, CT 06511}
\affiliation{Yale Quantum Institute, Yale University, New Haven, CT 06511}
\author{Leonid I. Glazman}
\email{leonid.glazman@yale.edu}
\affiliation{Department of Applied Physics, Yale University, New Haven, CT 06511}
\affiliation{Yale Quantum Institute, Yale University, New Haven, CT 06511}
\affiliation{Department of Physics, Yale University, New Haven, CT 06511}
\author{Robert J. Schoelkopf}
\email{robert.schoelkopf@yale.edu}
\affiliation{Department of Applied Physics, Yale University, New Haven, CT 06511}
\affiliation{Yale Quantum Institute, Yale University, New Haven, CT 06511}

\maketitle

\vspace*{-10mm}

\tableofcontents

\clearpage

\section{Experimental details}

\begin{figure}[h!]
    \centering
    \includegraphics[width=0.9\textwidth]{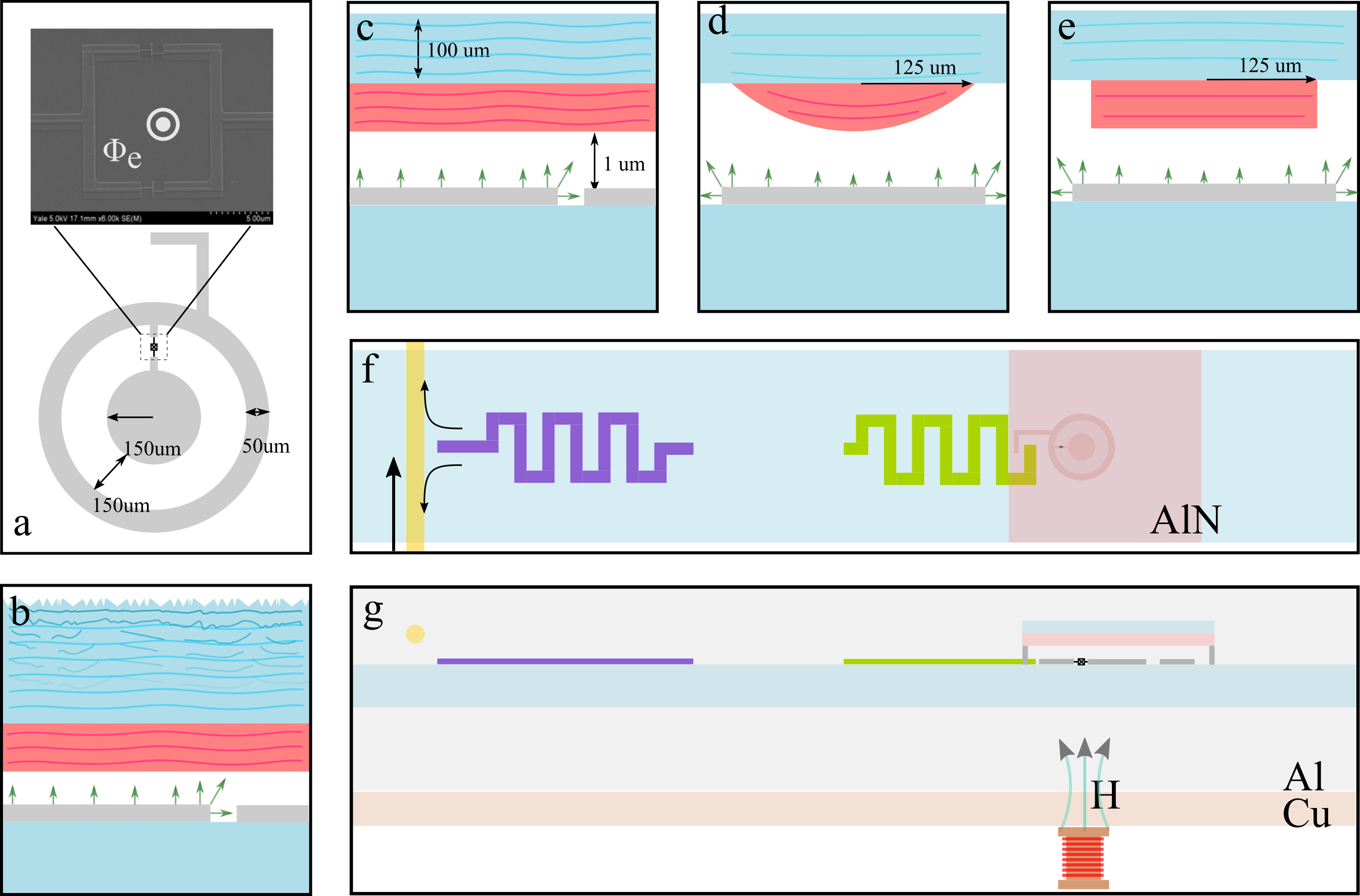}
    \caption{{\bf A schematic layout of the experimental setup.}
    (a)~A ringmon qubit: The qubit consists of an inner circular electrode, and an outer ring-shaped electrode. The electrodes are connected by a pair of parallel Josephson junctions, which form a SQUID loop. The loop is threaded by an external flux $\Phi_{\rm e}$ which controls the qubit frequency $\omega_0$.
    Panels (b, c, d, e) depict the four considered devices, which differ by the geometry of the acoustic resonator [the dimensions of the setup components in the $z$-directions are distorted for clarity]. (b)~A sapphire chip of thickness $430$~$\mu$m has a roughened backside, which leads to the diffuse scattering of the acoustic wave; the wave is excited by the electric field of the qubit via a piezoelectric effect in the $1$~$\mu$m thick AlN film.
    (c)~A qubit is coupled to a double-side polished sapphire chip of thickness $100$~$\mu$m.
    (d, e)~The dome and cylinder transducers are patterned on the surface of the resonator by selectively etching the AlN film.
    (f) A top-down view of the chip layout. From left to right are a Purcell filter (in purple), readout resonator (in green), and ringmon qubit (in grey). Stacked on top of the qubit is an HBAR consisting of an extended film of aluminum nitride on sapphire. The qubit is capacitively coupled to the readout resonator for dispersive readout. Microwave pulses are delivered to the system via a transmission line pin (yellow).
    (g) A side-cross section of the chip package. The body is made of aluminum, which superconducts, and the lid is made of copper, a normal metal that allows an external magnetic field to penetrate into the package. Flux threading the SQUID loop is used to tune the qubit's transition frequency. Aluminum spacers are deposited on the HBAR to create a nominal $1~\mu$m vacuum gap. This flip-chip assembly sits primarily in the superconducting region of the package.
}
    \label{fig:experimental-geometry}
\end{figure}

A concentric ring transmon~\cite{braumueller16}, aka `ringmon', is patterned adjacent to a meandering stripline resonator on a sapphire chip, as illustrated in Fig.~\ref{fig:experimental-geometry}. {The ringmon consists of two aluminium pads; an inner circular pad is surrounded by a ring-shaped outer one. The pads are connected via two parallel Josephson junctions, which form a SQUID loop. The junctions are nominally identical; they are made by a bridge-free double-angle evaporation process with a controlled oxidation. The Josephson inductance of the SQUID at zero flux is approximately $L_{J} = 7$nH.}

The ringmon has a finger-like extension for capacitive coupling to the readout resonator {[the readout resonator is depicted in green in Fig.~\ref{fig:experimental-geometry}(b, c)]}. The length and thickness of the finger are designed to {match the dispersive coupling strength $\chi_{\rm qb-ro}$ with the readout resonator's linewidth from coupling to the transmission line, $\kappa_{\rm ro}$}. The qubit is protected from environmental radiation by a Purcell filter {[purple in Fig.~\ref{fig:experimental-geometry}(b, c)]} adjacent to the waveguide pin  {[yellow in Fig.~\ref{fig:experimental-geometry}(b, c)]} used to address the qubit and readout modes.

Stacked above the qubit is a second chip of $c$-plane  {sapphire}, with a nominal thickness of $100\,\mu$m for the devices of Fig.~\ref{fig:experimental-geometry}(e,f,g) and $430\,\mu$m for the device of Fig.~\ref{fig:experimental-geometry}(d).  {The bottom surface of the chip has} a $1\mu$m thick film of highly-oriented $c$-plane  {aluminum nitride} ({deposited by} OEM Group). Note the residual stress in the film is typically between 100-200~MPa and is typically inhomogeneously distributed across the wafer. The gap  {between the bottom and the top sapphire chips} of $1\mu$m is formed by  {aluminum} spacers evaporated at the corners of the HBAR chip  {[see Fig.~\ref{fig:experimental-geometry}({c})]}. The two chips are held together by a micro volume of GE varnish, which also acts as the thermalization link and is far away from the qubit electrodes.\\

In two of our experimental samples {, schematically depicted in Fig.~\ref{fig:experimental-geometry}(f,g),} the piezoelectric film is shaped to form either a cylindrical disk or a dome transducer. The shaping is done by a two-step process, first by patterning a photoresist and second by reactive ion etching in a mixture of BCl$_{3}$/Cl$_{2}$/Ar, which selectively etches the aluminum nitride over the underlying sapphire. The dome shape requires an extra vapor-phase reflow step, which lightly melts the resist and forms a photoresist bubble held together by surface tension. The cylinder is measured to have a height of $1~
\mu$m and a diameter of $250~\mu$m; the remaining piezoelectric film is etched away from the surface. The dome is measured to have a height of $z_0 = 1~\mu$m and a radius of $125~\mu$m, which results in a radius of curvature {$\mathcal{R} = r^{2}/2z_0=7.8$~mm} 

The two-chip assembly is mounted in a package consisting of a long channel and a waveguiding pin, as illustrated in Fig.~\ref{fig:experimental-geometry}. 
The resonator-qubit assembly is hanger-coupled to the common feedline (shown in yellow in Fig.~\ref{fig:experimental-geometry}(b)), which is used to both drive the readout, qubit, and HBAR, and to collect photons leaking out of the readout for dispersive measurements. While the package bulk, including the channel, is made of 6061 aluminum, which superconducts at our operation temperature, the lid is made of OFHC copper to allow for the magnetic field from a coil magnet exterior to the package to penetrate into the channel and flux bias the qubit.

\clearpage
\section{Electro-mechanical coupling\label{sec:coupling}}
In this section, we derive an expression for the piezoelectric coupling between the qubit and the phonons in the resonator.  
An acoustic wave propagating in the resonator leads to a deformation of the AlN film. Due to the piezoelectric effect in AlN, the deformation gives rise to an electric polarization $\vec{P}(\bm{r})$:
\begin{equation}
    \vec{P}(\bm{r}) = \overleftrightarrow{e}\,\vec{s}(\bm{r}), \,
\end{equation}
where $\overleftrightarrow{e}$ is the piezoelectric tensor in the stress-charge form and $\vec{s}$ is the strain associated with the acoustic wave (we use the Voigt notations in which $\overleftrightarrow{e}$ is a $3 \times 6$ matrix and $\vec{s}$ is a six-component vector $[s_{xx},s_{yy},s_{zz},2s_{yz},2s_{xz}, 2s_{xy}]^{\rm T}$). 
The strain-induced polarization interacts with the electric field $\vec{E}$ produced by the qubit. The interaction is described~by
\begin{align}
{\cal H} = - \int \vec{E}(\bm{r})\cdot \vec{P}(\bm{r})\,{\rm d}V \equiv - \int \vec{\sigma}(\bm{r})\cdot \vec{s}(\bm{r})\,{\rm d}V. \label{eq:coupling_qph}
\end{align}
Here the integration is performed over the volume of the piezo-film; in the second equality we introduced $\vec{\sigma}(\bm{r}) = \overleftrightarrow{e}^{\rm T} \vec{E}(\bm{r})$, which can be interpreted as the mechanical stress induced in the film by the qubit's electric field. For the $c$-axis oriented AlN (in which the $c$-axis is aligned with the $z$-direction),
the piezoelectric tensor has five non-vanishing components \cite{auld,deJong2015}  
\begin{equation}
    \overleftrightarrow{e} = \begin{pmatrix}
        0 & 0 & 0 & 0 & e_{15} & 0\\
        0 & 0 & 0 & e_{15} & 0 & 0\\
        e_{31} & e_{31} & e_{33} & 0 & 0 & 0
    \end{pmatrix}.
\end{equation}
Thus, the stress $\vec{\sigma}$ can be represented as:
\begin{align}
        \vec{\sigma}(\bm{r}) = \bigl[ e_{31} E_z(\bm{r}),\, e_{31} E_z(\bm{r}),\, e_{33} E_z(\bm{r}),\, e_{15} E_y(\bm{r}),\, e_{15} E_x(\bm{r}),\, 0\bigr] ^{\textrm T}. \label{eq:sigma}
\end{align}

In all of the devices we consider, the qubit predominantly emits the acoustic waves in the $z$-direction (as result of the smallness of the acoustic wavelength in comparison with the size of the qubit's pads). Consequently, for the purpose of describing the acoustic radiation, we can leave only the dominant components of the strain tensor $s_{zz}$, $s_{xz}$, and $s_{yz}$ in the Hamiltonian \eqref{eq:coupling_qph}. Then, using Eq.~\eqref{eq:sigma}, we obtain:
\begin{equation}\label{eq:coupling_H}
    {\cal H} = {\cal H}_{l} + {\cal H}_{\rm sh},\quad\quad {\cal H}_{l} = -e_{33}  \int E_z(\bm{r}) s_{zz}(\bm{r})\,{\rm d}V,\quad\quad {\cal H}_{\rm sh} = -2e_{15}\int  \bigl(E_x(\bm{r}) s_{xz}(\bm{r}) + E_y(\bm{r}) s_{yz}(\bm{r})\bigr) \,{\rm d}V.
\end{equation}
Here ${\cal H}_{l}$ and ${\cal H}_{\rm sh}$ describe the interaction of the qubit with longitudinal and shear waves, respectively. We note that $E_z$ is the largest component of the in-film electric field while $e_{33}$ is the dominant component of the piezoelectric tensor (see Sec.~\ref{sec:mat-params}). This means that the qubit couples the strongest to the longitudinal waves. Thus, we focus on ${\cal H}_{l}$ in the remainder of the section, and use it to find the coupling strength $g_\nu$ of the qubit to a longitudinal phonon mode $\nu$.

To find $g_\nu$, we expand the strain field entering ${\cal H}_l$ into the phonon modes of the resonator.
A convenient way to do this is to first relate the strain to the crystalline lattice displacement $\vec{u}(\bm{r})$ via the definition $s_{ij} = \frac{1}{2}\bigl(\partial u_i / \partial r_j + \partial u_j / \partial r_i\bigr)$.
The displacement field can be expanded into the phonon modes labelled by an index $\nu$ as
\begin{equation}
    \vec{u}(\bm{r})	=\sum_{\nu}\sqrt{\frac{\hbar}{2\rho\omega_{\nu}}}\left(a_{\nu}\vec{u}_{\nu}(\bm{r})+a_{\nu}^{\dagger}\vec{u}_{\nu}^{\star}(\bm{r})\right).\label{eq:displacement_expansion}
\end{equation}
Here $a_{\nu}$ is the annihilation operator for a phonon in a given mode $\nu$, $\omega_{\nu}$ is the frequency of this mode, and $\vec{u}_{\nu}$ is the corresponding displacement field. We assume the following normalization condition: $\int \vec{u}_{\nu}(\bm{r})\cdot\vec{u}_{\nu^{\prime}}^{\star}(\bm{r})\,{\rm d} V=\delta_{\nu\nu^{\prime}}$. The combination $\vec{u}_{\nu}(\bm{r})\sqrt{\hbar/2\rho\omega_{\nu}}$ is the zero-point fluctuation of displacement associated with the mode $\nu$; $\rho$ is the density of the acoustic medium.
Substituting decomposition \eqref{eq:displacement_expansion} into ${\cal H}_{l}$, we obtain
\begin{equation}\label{eq:H_l_modes}
{\cal H}_l = -e_{33}\sum_\nu \sqrt{\frac{\hbar}{2\rho\omega_{\nu}}} \Bigl(a_\nu \int E_z(\bm{r}) \partial_z u_{\nu,z}(\bm{r})\,{\rm d}V + a_\nu^\dagger  \int E_z(\bm{r}) \partial_z u^\star_{\nu,z}(\bm{r})\,{\rm d}V \Bigr),
\end{equation}
where $\partial_z \equiv \frac{\partial}{\partial z}$.

Finally, we note that $E_z$ is in fact an operator acting on the qubit degree of freedom. To make this explicit, we promote $E_z\rightarrow E_z\sigma_{x}$. Upon making this replacement in Eq.~\eqref{eq:H_l_modes} and leaving only the resonant terms\footnote{The off-resonant terms $a_{\nu}^\dagger \sigma^+$ and $a_{\nu}^\dagger \sigma^-$  lead to renormalization of the system parameters; e.g., they produce (a contribution to) the Lamb shift of the qubit frequency. Focusing on the phonon radiation by a qubit, we dispense with such effects.} $a_{\nu}\sigma^{+}$ and $a_{\nu}^{\dagger}\sigma^{-}$, we arrive to
\begin{equation}\label{eq:coupling_general}
    {\cal H}_l = \hbar \sum_\nu \bigl(g_\nu\,a_\nu \sigma^+ + g_\nu^\star\,a_\nu^\dagger\, \sigma^-\bigr),\quad\quad\text{where}\quad\quad g_\nu = -\frac{e_{33}}{\sqrt{2\hbar \rho \omega_\nu}}\int E_z(\bm{r}) \partial_z u_{\nu,z}(\bm{r})\,{\rm d}V.
\end{equation}
The parameter $g_\nu$ determines the coupling strength between the qubit and the acoustic mode $\nu$.

\subsubsection*{Coupling to the modes in the double-side polished acoustic resonator}
Let us now apply Eq.~\eqref{eq:coupling_general} to find the coupling of the qubit to phonon modes in the flat resonator of thickness $\thickness$ [see Fig.~\ref{fig:experimental-geometry}(d)].
Two parallel, polished surfaces of such a resonator result in the formation of the standing waves in it.
The acoustic modes can be labelled by the standing wave overtone number $n$ and the in-plane wave-vector $\bm{k}_{\perp}$. The displacement field for a longitudinal wave with $k_{\perp}\ll k_{z} \equiv \pi n /\thickness$ is given by
\begin{equation}
\vec{u}_{n,k_{\perp}}(\bm{r})	=\sqrt{\frac{2}{V_{{\rm cav}}}}{\rm cos}\left(\frac{\pi nz}{\thickness}\right)e^{i\bm{k}_{\perp}\cdot\bm{r}_{\perp}}\hat{z},\label{eq:u_phonon_mode}
\end{equation}
where $\hat{z}$ is the unit vector in the $z$-direction and $V_{\rm cav}$ is the total volume of the cavity. We assume that the bottom surface of resonator is at $z=0$; the displacement has an anti-node there, reflecting the free surface boundary condition (i.e., the vanishing of the mechanical stress at the surface). The factor of $\sqrt{2}$ stems from the normalization condition. The frequency of the mode $\{n,\bm{k}_{\perp}\}$ is given by \begin{equation}\label{eq:spectrum}
    \omega_{n}(\bm{k}_{\perp}) = \sqrt{\left(\frac{\pi nv_{l}}{\thickness}\right)^{2}+v_{\perp}^{2}k_{\perp}^{2}}.
\end{equation}
Here $v_{l}$ and $v_\perp$ are the two velocities characterizing the dispersion of waves. These velocities are related to the elastic stiffness tensor $c_{ij}$ of the acoustic medium. Specifically, the longitudinal wave velocity is given by $v_l^2 = c_{33} / \rho$, whereas $v_\perp^2 = [c_{44} + (c_{13} + c_{44})^2 / (c_{33} - c_{44})]/\rho$ [see Sec.~\ref{sec:mat-params} for the values of $c_{ij}$ for sapphire]. 

Using Eq.~\eqref{eq:u_phonon_mode} in Eq.~\eqref{eq:coupling_general}, we find for the qubit-phonon coupling:
\begin{equation}\label{eq:H_l_final}
    {\cal H}_l = \hbar\sum_{n,\bm{k}_{\perp}}\left(g_{n\bm{k}_{\perp}}a_{n\bm{k}_{\perp}}\sigma^{+}+g_{n\bm{k}_{\perp}}^{\star}a_{n\bm{k}_{\perp}}^{\dagger}\sigma^{-}\right),\quad\quad g_{n\bm{k}_{\perp}}=\frac{2e_{33}\hat{E}_{z}(\bm{k}_{\perp})}{\sqrt{\hbar\rho\omega_{n}(k_{\perp})V_{{\rm cav}}}}{\rm sin}^{2}\Bigl(\frac{\pi n \thickness_{p}}{2\thickness}\Bigr),
\end{equation}
where $\hat{E}_z(\bm{k}_\perp) = \int d^2\bm{r}_\perp e^{-i\bm{k}_\perp \cdot \bm{r}_\perp} E_z(\bm{r}_\perp)$ is a 2D Fourier transform of the electric field (since the piezoelectric film employed in the experiments is thin compared to the qubit's size, $\sim 1\,{\rm \mu m} \ll 350\,{\rm \mu m}$, we dispense with the weak dependence of the in-film electric field on $z$). The factor $2\,{\rm sin}^2 (\pi n \thickness_p / 2 \thickness) $ originates from the integration of the strain mode over the thickness $\thickness_p$ of the piezoelectric film. Expressions similar to Eq.~\eqref{eq:H_l_final} can be obtained straightforwardly for the coupling of the qubit to the shear phonons.

\clearpage
\section{Calculation of acoustic spontaneous emission into a rough-backside resonator}
Here, we consider a qubit coupled to a rough-backside resonator, and find a contribution to its relaxation rate associated with the emission of acoustic waves.
As explained in Sec.~IIA of the main text, the emission of waves into a rough-backside resonator happens with the same rate as it does into an acoustic free space.
For the purpose of finding the rate, we focus on the latter, simpler configuration. 

A convenient way to describe the acoustic free space is to consider a flat resonator with thickness $\thickness \rightarrow \infty$. The modes in the flat resonator are indexed by $\{ n , \bm{k}_{\perp} \}$. The integer $n$ here determines the discrete values of the $z$-component of the wavevector, $k_z(n) = \pi n / b$.
To find the rate of the free-space longitudinal wave emission, we apply Fermi's Golden Rule:
\begin{equation}\label{eq:FGR_emission}
    \gamma_{{\rm fs},l}(\omega_0) = 2\pi \sum_{n, \bm{k}_\perp} |g_{n\bm{k}_\perp}|^2\delta (\omega_0 - \omega_n(\bm{k}_\perp)).
\end{equation}
The discretization step $\Delta k_z = \pi / b$ vanishes in the limit $b \rightarrow \infty$. Therefore, one can treat $k_z$ as a continuous variable, which allows us to replace $\sum_{n, \bm{k}_\perp} = V_{\rm cav}\int_0^\infty \frac{dk_z}{\pi}\int \frac{d^2\bm{k}_\perp}{(2\pi)^2}$ in Eq.~\eqref{eq:FGR_emission}. Then, with the help of Eqs.~\eqref{eq:spectrum} and \eqref{eq:H_l_final}, we find:
\begin{align}
\gamma_{{\rm fs},l}(\omega_0) = &\,\frac{2\pi}{\hbar}\left[\frac{4e_{33}^{2}}{\rho\omega_0}{\rm sin}^{4}\left(\frac{\omega_0\thickness_{p}}{2v_{l}}\right)\right]\cdot\int_0^\infty \frac{dk_z}{\pi}\int\frac{d^{2}\bm{k}_{\perp}}{(2\pi)^{2}}|\hat{E}_{z}(\bm{k}_{\perp})|^{2}\delta\bigl(\omega_0-\sqrt{v_l^2 k_z^2 + v_{\perp}^2 k_\perp^2}\bigr) \label{eq:fs_long}\\
=&\,\frac{2\pi}{\hbar}\left[\frac{4e_{33}^{2}}{\pi v_l \rho\omega_0}{\rm sin}^{4}\left(\frac{\omega_0\thickness_{p}}{2v_{l}}\right)\right]\cdot\int\frac{d^{2}\bm{k}_{\perp}}{(2\pi)^{2}}|\hat{E}_{z}(\bm{k}_{\perp})|^{2} \frac{\omega_0}{\sqrt{\omega_0^2 - v_{\perp}^2 k_\perp^2}}.
\notag
\end{align}
The dependence of $|{\hat E}_{z}(\bm{k}_{\perp})|^{2}$ on $k_\perp$ is determined by the geometry of the transmon: $|{\hat E}_{z}(\bm{k}_{\perp})|^{2}$ peaks at $k_{\perp}\sim k^{\rm pk}_{\perp} \equiv 2\pi / a$, with $a \approx 350\,{\rm \mu m}$ being the transmon's spatial footprint, and quickly decays at $k_\perp \gtrsim k_\perp^{\rm pk}$.
This means that the qubit predominantly emits phonons with $k_\perp \sim k_\perp^{\rm pk} \ll 2\pi / \lambda_{\rm ac}$, where $\lambda_{\rm ac} \sim 2\,{\rm \mu m}$ is the acoustic wavelength at the qubit frequency $\omega_0 \sim 2\pi \cdot 6\,{\rm GHz}$. The latter condition allows us to neglect $k_\perp$ under the square root in Eq.~\eqref{eq:fs_long}. Then, we obtain:
\begin{align}
\gamma_{{\rm fs},l}(\omega_0) 
=&\,\frac{2\pi}{\hbar}\left[\frac{4e_{33}^{2}}{\pi v_l \rho\omega_0}{\rm sin}^{4}\left(\frac{\omega_0\thickness_{p}}{2v_{l}}\right)\right]\cdot\int d^{2}\bm{r}_{\perp}E^2_{z}(\bm{r}_{\perp}).\label{eq:fs_long_final}
\end{align}
Here we used the Parceval's identity, $\int\frac{d^{2}\bm{k}_{\perp}}{(2\pi)^{2}}|\hat{E}_{z}(\bm{k}_{\perp})|^{2} \equiv\int d^{2}\bm{r}_{\perp}E_{z}^{2}(\bm{r}_{\perp})$ to express the integral over the wavevectors as a real-space integral. 

To find the rate of the shear waves emission, we start with the coupling Hamiltonian ${\cal H}_{\rm sh}$ (see Eq.~\eqref{eq:coupling_qph}) and repeat the steps leading to Eq.~\eqref{eq:fs_long_final}. In full analogy to $\gamma_{l}$ we obtain
\begin{align}
\gamma_{{\rm fs},{\rm sh}, \alpha}(\omega_0) & =\frac{2\pi}{\hbar}\left[\frac{4e_{15}^2}{\pi v_{\rm sh} \rho\omega_0} {\rm sin}^{4}\left(\frac{\omega_0\thickness_{p}}{2v_{\rm sh}}\right)\right]\cdot\int d^{2}\bm{r}_{\perp}E_{\alpha}^{2}(\bm{r}_{\perp}),\label{eq:gamma_sh}
\end{align}
where $\alpha = x, y$ corresponds to the two possible polarizations of the shear waves.

\begin{table}[b]
 \begin{tabular}{|c|c|c|c|c|c|c|c|c|} 
 \hline
 $\rho,\,{\rm kg/m^3}$ & $v_{l},\,{\rm km / s}$ & $v_{\rm sh},\,{\rm km / s}$ & $e_{33},\,{\rm C / m^2}$ & $e_{15},\,{\rm C / m^2}$  & $\thickness_p,\,{\rm \mu m}$ & $\int d^2{\bm r}_\perp E_x^2(\bm{r}_\perp),\,{\rm V^2}$ & $\int d^2{\bm r}_\perp E_y^2(\bm{r}_\perp),\,{\rm V^2}$ & $\int d^2{\bm r}_\perp E_z^2(\bm{r}_\perp),\,{\rm V^2}$\\ [0.5ex] 
 \hline
 3980 & 11.2 & 6.1 & 1.40 & 0.40 & 1.0 & $2.33 \cdot 10^{-10}$ & $2.17 \cdot 10^{-10}$ & $5.75 \cdot 10^{-10}$ \\ [0.5ex]
 \hline
 \end{tabular}
 \caption{Here, $\rho$ is the density of sapphire; $v_l$ and $v_{\rm sh}$ are the velocities of longitudinal and shear waves propagating in the sapphire crystal along the $c$-axis (which is aligned with the $z$-axis of the device); the velocities are given by $v_l = \sqrt{c_{33}/\rho}$ and $v_{\rm sh} = \sqrt{c_{44}/\rho}$, where $c_{ij}$ tensor is given in Sec.~\ref{sec:mat-params}; $e_{33}$ and $e_{15}$ are the components of the piezoelectric tensor for the polycrystalline AlN film (see Sec.~\ref{sec:mat-params} for the discussion of values being used),
 $\thickness_p$ is the thickness of the AlN film, and $E_{x,y,z}(\bm{r}_\perp)$ are the components of the qubit's electric field in the AlN film.\label{tab:parameters}}
\end{table}

\begin{figure}[t]
    \centering
    \includegraphics[width=0.8\textwidth]{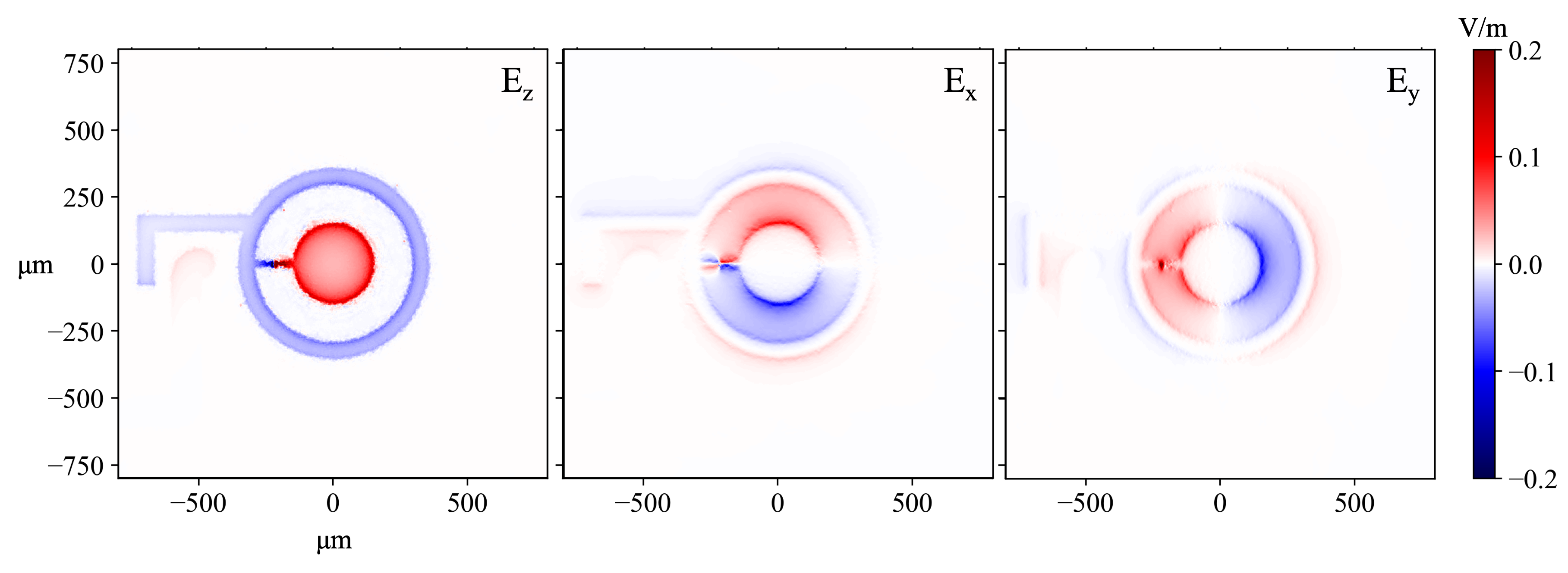}
    \caption{\textbf{The electric field in the AlN film.} The electric field distributions are calculated from the results of an HFSS eigenmode simulation with a single photon of energy.
    }
    \label{fig:app-e-fields-hfss}
\end{figure}

The total rate of the spontaneous emission is given by $\gamma_{\rm fs} = \gamma_{{\rm fs},l} + \gamma_{{\rm fs},{\rm sh}, x} + \gamma_{{\rm fs},{\rm sh}, y}$.
The material and device parameters needed to estimate the three contributions to $\gamma$ are listed in Table \ref{tab:parameters}.
The acoustic wave velocities in Table \ref{tab:parameters} correspond to the sapphire crystal; in the estimates, we dispense with the difference in velocities between sapphire and AlN of less than $5\%$. 
We have not measured the piezoelectric constants of our AlN film independently, so we use the previously reported values  \cite{muralt1999, muralt2001, guy99, muensit99} in our estimates. 
To find the electric field in the AlN film, we employ the high frequency simulation software (HFSS). The results of the HFSS simulation for $E_{x,y,z}({\bm r}_\perp)$ are shown in Fig.~\ref{fig:app-e-fields-hfss}. The simulated single-photon field distributions yield the numbers presented in the last three columns of Table~\ref{tab:parameters}.

Using Eqs.~\eqref{eq:fs_long_final}, \eqref{eq:gamma_sh} and Table~\ref{tab:parameters} we find at the qubit frequency of $\omega_0 / 2\pi = 2\pi \cdot 6.69\,{\rm GHz}$:
\begin{subequations}
\begin{align}
    \gamma_{{\rm fs},l} / 2\pi =& \,5.9\,{\rm MHz},\\
    \gamma_{{\rm fs},{\rm sh}, x} / 2\pi =& \,7.0\,{\rm kHz},\\
    \gamma_{{\rm fs},{\rm sh}, y} / 2\pi =& \,6.6\,{\rm kHz}.
\end{align}
\end{subequations}
Remarkably, the rate of the shear waves emission is smaller than the rate of the longitudinal waves emission by nearly three orders of magnitude. 
In part, this smallness originates from the smallness of $e_{15}$ in comparison with $e_{33}$, and that of $E_{x,y}$ in comparison with $E_z$, see Table~\ref{tab:parameters}.
The shear waves emission is further diminished by the sine factor in Eq.~\eqref{eq:gamma_sh}.
Indeed, the wavelength of the shear wave at $\omega_0/2\pi = 6.69\,{\rm GHz}$ is close to the width of the film $\thickness_p$, which results in ${\rm sin}^4 (\omega_0 \thickness_p / 2 v_{\rm sh}) = 1.6\cdot 10^{-2} \ll 1$.
By contrast, $\lambda_{{\rm ac},l} \approx 2\thickness_p$ for a longitudinal wave and so ${\rm sin}^4 (\omega_0 \thickness_p / 2 v_{l}) = 0.8 \sim 1$. 
Since $\gamma_{{\rm fs},\rm sh} \ll \gamma_{{\rm fs},l}$, we disregard the emission of the shear waves in Sec.~IIA of the main text. Specifically, in Eq.~(1) we approximate $\gamma_{\rm fs} \simeq \gamma_{{\rm fs},l}$ and use Eq.~\eqref{eq:fs_long_final}.

\newpage

\section{Dynamics of the qubit coupled to a double-side polished chip\label{sec:double-side-polished}}
In this section, we describe the dynamics of the qubit coupled to a double-side polished acoustic resonator.
An initially excited qubit may lose its energy by emitting a phonon. 
Due to the formation of a standing wave in the double-side polished chip, the emitted phonon may be reabsorbed by the qubit at a later time; in this case vacuum Rabi oscillations emerge. 
Another possibility is that the emitted phonon does not return to the qubit; then the qubit relaxes its energy irreversibly. 
Below, we establish the conditions that determine which of the two regimes is realized.
We use the results of this section to derive the estimates for $g$ and $\gamma_{\rm R}$ presented in Sec.~IIB of the main text.

We start with the Hamiltonian of the qubit coupled to the acoustic modes in the chip:
\begin{equation}\label{eq:Hamiltonian}
    {\cal H} / \hbar = \frac{\omega_0}{2} (\sigma_z + 1) + \sum_{n, \bm{k}_\perp} \omega_n(\bm{k}_\perp) a_{n\bm{k}_\perp}^\dagger a_{n\bm{k}_\perp} + \sum_{n,\bm{k}_{\perp}}\left(g_{n\bm{k}_{\perp}}a_{n\bm{k}_{\perp}}\sigma^{+}+g_{n\bm{k}_{\perp}}^{\star}a_{n\bm{k}_{\perp}}^{\dagger}\sigma^{-}\right).
\end{equation}
The first and the second terms describe the qubit and the acoustic modes, respectively.
The acoustic modes are labeled by the standing wave overtone number $n$ and the in-plane wavevector $\bm{k}_\perp$. Throughout the section, we focus on longitudinal waves, whose coupling to the qubit is the strongest, and leave the wave polarization label implicit.
The experimentally relevant regime corresponds to the interaction of the qubit with the high-overtone mode, $n \gg 1$.
For $k_\perp \ll 2\pi / \lambda_{\rm ac}$ ($\lambda_{\rm ac}$ is the acoustic wavelength at the qubit frequency $\omega_0$), the frequency of such a mode can be approximated by
\begin{equation}
    \omega_n(\bm{k}_\perp) \simeq \omega_n + \frac{v_\perp^2 k_\perp^2}{2\omega_n},
\end{equation}
where $\omega_n = \pi n v_l / b$ is the standing wave frequency. The third term in Eq.~\eqref{eq:Hamiltonian} describes the coupling between the qubit and the longitudinal phonons. It was derived in Sec.~\ref{sec:coupling}, see Eq.~\eqref{eq:H_l_final} for the coupling strength $g_{n\bm{k}_\perp}$. 

To elucidate the qubit dynamics, we assume that the system is initialized in a state $\left|e, 0_\mathrm{ph}\right\rangle$, which corresponds to an excited qubit ($e$) on the background of the phonon vacuum. In the course of evolution described by the Hamiltonian \eqref{eq:Hamiltonian}, this state mixes with a superposition of states $a_{n\bm{k}_\perp}^\dag\left|g, 0_\mathrm{ph}\right\rangle$ corresponding to the relaxed qubit ($g$) and an emitted phonon. The respective wavefunction has the form:
\begin{equation}
    |\psi(t)\rangle = A_e(t) |e, 0_{\rm ph}\rangle + \sum_{n, \bm{k}_\perp} A_{g, n\bm{k}_\perp}(t)\, a_{n\bm{k}_{\perp}}^\dagger |g, 0_{\rm ph}\rangle.
\end{equation}
It satisfies the time-dependent Schr\"odinger equation $i\hbar \partial_t |\psi(t)\rangle = {\cal H} |\psi(t)\rangle$, which reduces to a system of coupled equations for amplitudes $A_e(t)$ and $A_{g,n\bm{k}_\perp}(t)$: 
\begin{subequations}\label{eq:A_system}
\begin{align}
    i \frac{dA_e(t)}{dt} &= \omega_0 A_e(t) + \sum_{n, \bm{k}_\perp} g_{n\bm{k}_\perp} A_{g, n\bm{k}_\perp}(t) + i \delta(t),\label{eq:A_system_1}\\
    i \frac{dA_{g, n\bm{k}_\perp}(t)}{dt} &=  \omega_n(\bm{k}_\perp) A_{g, n\bm{k}_\perp}(t) + g_{n\bm{k}_\perp}^\star A_e(t).
\end{align}
\end{subequations}
We introduced the delta-function in Eq.~\eqref{eq:A_system_1} to account for the initial condition $A_{e} (t = 0) = 1$. Keeping in mind the focus on the emission of the acoustic waves, we disregard other possible mechanisms of the qubit relaxation in Eq.~\eqref{eq:A_system}. Furthermore, we assume a high-Q acoustic resonator and thus treat phonons as the conservative subsystem.

In the frequency-domain, Eq.~\eqref{eq:A_system} boils down to a system of algebraic equations:
\begin{subequations}
\begin{align}
    \omega A_e(\omega) &= \omega_0 A_e(\omega) + \sum_{n, \bm{k}_\perp} g_{n\bm{k}_\perp} A_{g, n\bm{k}_\perp}(\omega) + i,\\
    \omega A_{g, n\bm{k}_\perp}(\omega) &=  \omega_n(\bm{k}_\perp) A_{g, n\bm{k}_\perp}(\omega) + g_{n\bm{k}_\perp}^\star A_e(\omega).
\end{align}
\end{subequations}
Solving this system for $A_{e}(\omega)$ and converting the result back into the time-domain, we find:
\begin{equation}\label{eq:solution_for_A}
    A_e(t) = i\int \frac{d\omega}{2\pi} \frac{e^{-i\omega t}}{\omega + i0^+ - \omega_0 - \Sigma(\omega)},\quad\quad \text{where}\quad\quad\Sigma(\omega) = \sum_{n \bm{k}_\perp} \frac{|g_{n\bm{k}_\perp}|^2}{\omega + i0^+ - \omega_{n}(\bm{k}_\perp)}.
\end{equation}
An infinitesimally small positive number $0^+$ in the denominators ensures causality. The ``self-energy'' $\Sigma(\omega)$ describes the qubit-phonon interaction; it contains all of the information on the phonon subsystem relevant for the dynamics of $A_e(t)$.

The self-energy $\Sigma(\omega)$ can be found under a set of simplifying assumptions. First, we shall assume that the qubit frequency $\omega_0$ is close to a resonance with the standing wave frequency $\omega_n = \pi n v_l / \thickness$ specified by an overtone number $n$. 
For sufficiently weak interaction, this allows us to dispense with the terms whose overtone numbers $n^\prime \neq n$ in the definition of $\Sigma(\omega)$ (the applicability of this condition is discussed at the end of the section).
Using this simplification together with expression \eqref{eq:H_l_final} for $g_{n\bm{k}_\perp}$, and also changing the summation over $\bm{k}_\perp$ to integration, we find for $\Sigma(\omega)$:
\begin{equation}\label{eq:self-energy}
    \Sigma(\omega) = \frac{4 e_{33}^2}{\hbar \rho \omega_n b} {\rm sin}^4\Bigl(\frac{\pi n b_p}{2 b}\Bigr) \int \frac{d^2 \bm{k}_\perp}{(2\pi)^2} \frac{|\hat{E}_z(\bm{k}_\perp)|^2}{\omega + i0 - \omega_n(\bm{k}_\perp)}.
\end{equation}
To evaluate the integral over the in-plane wavevectors, we need to specify the Fourier-transform of the electric field $\hat{E}_z(\bm{k}_\perp)$.
This is a formidable task for a generic field distribution (e.g., the one depicted in Fig.~\ref{fig:app-e-fields-hfss}). To proceed analytically, we shall consider a simplified geometry of the system. Specifically, we will assume that only one circular pad of the transmon (of radius $a$) is piezoelectrically coupled to the chip.
We will also treat the electric field $E_z$ as being spatially uniform across the pad's area.
We expect that such a simplified model gives rise to all of the essential features in the qubit dynamics.
The particular quantitative predictions can be used as the order-of-magnitude estimates for comparison with the experimental data.
Under the above assumptions, the Fourier-transform of the electric field is given by
\begin{equation}\label{eq:Ez_Fraunhofer}
    \hat{E_z}(\bm{k}_\perp) = E_z \pi a^2\cdot \Bigl[\frac{2 J_1 (k_\perp a)}{k_\perp a}\Bigr],
\end{equation}
where $J_1(z)$ is the Bessel function. 
Note that the dependence of $\hat{E_z}(\bm{k}_\perp)$ on $k_\perp$ (and, accordingly, that of the coupling strength $g_{n\bm{k}_\perp}$) is oscillating; it demonstrates a conventional Fraunhofer diffraction pattern.
Substituting expression \eqref{eq:Ez_Fraunhofer} into Eq.~\eqref{eq:self-energy}, and computing the integral over $\bm{k}_\perp$ we find:
\begin{equation}\label{eq:self-energy_result}
    \Sigma(\omega) =
    \frac{g^2}{\omega - \omega_n}\times\begin{cases}
    \Bigl[ 1 - 2 I_1 \Bigl(\pi \sqrt{ \frac{\omega_n - \omega}{\omega_{\rm diff}}}\Bigr) K_1 \Bigl(\pi \sqrt{\frac{\omega_n - \omega}{\omega_{\rm diff}}}\Bigr)\Bigr], \quad\quad \omega < \omega_n,\vspace{0.2cm}\\
    \Bigl[ 1 + \pi J_1 \Bigl(\pi \sqrt{\frac{\omega - \omega_n}{\omega_{\rm diff}}}\Bigr) Y_1 \Bigl(\pi \sqrt{\frac{\omega - \omega_n}{\omega_{\rm diff}}}\Bigr) - i \pi J_1^2 \Bigl(\pi \sqrt{\frac{\omega - \omega_n}{\omega_{\rm diff}}}\Bigr)\Bigr], \quad\quad \omega > \omega_n.
    \end{cases}
\end{equation}
Here $I_1(z)$ and $K_1(z)$ are the modified Bessel functions of the first and second kind, respectively, and $Y_1(z)$ is the Neumann function. Parameter
\begin{equation}\label{eq:omega_diff}
\omega_{\rm diff} = v_\perp^2 \pi^2 /(2\omega_n a^2)
\end{equation}
has a meaning of the frequency separation (up to a number $\sim 1$) between the phonon modes corresponding to the zeroth and the first diffraction maxima [cf.~Eq.~\eqref{eq:Ez_Fraunhofer}]. Finally, in Eq.~\eqref{eq:self-energy_result} we introduced
\begin{equation}\label{eq:g_standing_wave}
    g = \frac{2 e_{33}}{\sqrt{\hbar \rho \omega_n V_{\rm mode}}} {\rm sin}^2 \Bigr(\frac{\pi n b_p}{2b}\Bigr) E_z \pi a^2.
\end{equation}
This parameter has the meaning of the coupling strength between the qubit and the standing wave confined in the volume $V_{\rm mode} = b\hspace{0.05cm} \pi a^2$ above the transmon pad. As we will see momentarily, the qualitative character of the qubit dynamics---including the emergence of vacuum Rabi oscillations---depends on the relation between $g$ and the diffraction frequency scale $\omega_{\rm diff}$.

From the dependence of $\Sigma(\omega)$ 
on $\omega - \omega_n$ we see that the coupling of the qubit to the acoustic medium decreases with the increase of $\omega_0$ above the threshold $\omega_0 = \omega_n$.
A sufficiently strong detuning $\omega_0 - \omega_n$ brings the system into a perturbative regime of the qubit-phonon interaction.
Assuming that the perturbation theory is valid (the precise condition will be specified shortly), it is sufficient to account for a shift of the pole in Eq.~\eqref{eq:solution_for_A}, $\omega_0 \rightarrow \omega^\prime_q - i \gamma(\omega_0) / 2 $, where $\omega_0^\prime \approx \omega_0 + \mathrm{Re}\,\Sigma(\omega_0)$ and $\gamma(\omega_0)  \approx -2\,\mathrm{Im}\,\Sigma(\omega_0)$. 
Then, we find:
\begin{equation}\label{eq:pert}
    A_e(t) \approx e^{-i\omega_0^\prime t} e^{-\gamma(\omega_0) t / 2},\quad\quad\text{where}\quad\quad \gamma(\omega_0) = 2\pi g^2\cdot \frac{J_1^2 \Bigl(\pi \sqrt{\frac{\omega_0 - \omega_n}{\omega_{\rm diff}}}\Bigr)}{\omega_0 - \omega_n}.
\end{equation}

Equation \eqref{eq:pert} describes an exponential decay of the qubit excited state population $|A_e(t)|^2$, with a decay constant $\gamma(\omega_0)$.
Expectedly, this prediction agrees with the result of the Fermi's Golden Rule application to the problem of qubit coupled to the phonon continuum. 
For $\omega_0 - \omega_n \gg \omega_{\rm diff}$, we can approximate the decay rate by
\begin{equation}\label{eq:decay}
     \gamma(\omega_0) = \overline{\gamma}(\omega_0)\,{\rm sin}^2 \Bigl(\pi \sqrt{\frac{\omega_0 - \omega_n}{\omega_{\rm diff}}} - \frac{\pi}{4}\Bigr),\quad\text{with the envelope function}\quad\overline{\gamma} (\omega_0) = \frac{4}{\pi}\frac{g^2}{\omega_{\rm diff}\,\bigl[({\omega}_q - \omega_n)/\omega_{\rm diff}\bigr]^{3/2}}.
\end{equation}
The oscillatory factor in $\gamma(\omega_0)$ results from the diffraction of phonons on the sharply varying electric field profile of the qubit; the spacing between two subsequent diffraction maxima in $\gamma(\omega_0)$ is $\Delta\omega_0 \sim  \omega_{\rm diff} \sqrt{1 + (\omega_0 - \omega_n) / \omega_{\rm diff}}$. Note that the envelope function decreases above the threshold frequency $\omega_n$, $\overline{\gamma} (\omega_0) \propto (\omega_0 - \omega_n)^{-3/2}$; this reflects the mentioned weakening of the qubit-phonon coupling with the increase of $\omega_0 - \omega_n$. 

We now establish the applicability conditions for the perturbative treatment.
By inspecting the denominator of Eq.~\eqref{eq:solution_for_A} for $A_e(t)$,
we determine that Eq.~\eqref{eq:pert} is valid as long as the spacing between the consequent diffraction maxima exceeds the typical decay decrement in the vicinity of $\omega_0$,
\begin{equation}\label{eq:cond}
    \overline{\gamma}(\omega_0) \ll \omega_{\rm diff} \sqrt{1 + (\omega_0 - \omega_n) / \omega_{\rm diff}}.
\end{equation}
If when decreasing $\omega_0$ condition \eqref{eq:cond} remains satisfied down to $\omega_0 - \omega_n \sim \omega_{\rm diff}$, then the qubit
undergoes a featureless relaxation regardless of its frequency; in particular, no Rabi vacuum oscillations emerge. This weak-coupling regime is realized for $g \ll \omega_{\rm diff}$, as can be verified using Eqs.~\eqref{eq:decay} and \eqref{eq:cond}.

The time-dependence of $A_e(t)$ is more intricate if the coupling $g$ exceeds the diffraction linewidth $\omega_{\rm diff}$;
then, the condition \eqref{eq:cond} breaks down for qubit frequency in vicinity of the threshold, $0<\omega_0 - \omega_n \lesssim g$. 
We expect well-developed vacuum Rabi oscillations to occur in this range of $\omega_0$.
To demonstrate the occurrence of Rabi oscillations, we derive an approximate expression for the probability $|A_e(t)|^2$.
For simplicity, we first concentrate on the early-time dynamics of the qubit. 
At $t \ll 1 / \gamma_{\rm R}$, the second and third terms in the square brackets of Eq.~\eqref{eq:self-energy_result} have a negligible influence on the evolution of $A_e(t)$ and can be disregarded (the estimate for $\gamma_{\rm R}$ will follow shortly). After dispensing with these terms, it is straightforward to find poles of the integrand in Eq.~\eqref{eq:solution_for_A}; it is these poles that determine the evolution of $A_e(t)$. By setting $\omega - \omega_0 - \Sigma(\omega,\nu) = 0$, we find two solutions for $\omega$,
\begin{equation}\label{eq:omega_pm}
    \omega_{\pm} = \frac{\omega_0 + \omega_n}{2} \pm \sqrt{g^2 + \frac{(\omega_0 - \omega_n)^2}{4}}\,,
\end{equation}
where $\omega_+ > \omega_n$ and $\omega_- < \omega_n$. Integral over $\omega$ in Eq.~\eqref{eq:solution_for_A} can be computed by summing up the residues at $\omega = \omega_{\pm}$. We obtain
\begin{equation}\label{eq:ampl_oscill}
    A_e(t) \approx \frac{\omega_+ - \omega_n}{\omega_+ - \omega_-} e^{-i\omega_+ t} -\frac{\omega_- - \omega_n}{\omega_+ - \omega_-} e^{-i\omega_- t}.
\end{equation}
From this equation, we find for the excited state population:
\begin{equation}\label{eq:solution}
    |A_e(t)|^2 \approx 1 - \frac{4g^2}{\Omega_{\rm R}^2} {\rm sin}^2 \Bigl[\frac{\Omega_{\rm R} t}{2}\Bigr],\quad \quad \Omega_{\rm R} = \sqrt{4g^2 + {(\omega_0 - \omega_n)^2}}\,.
\end{equation}
This expression highlights the oscillations of the population with time; these are the vacuum Rabi oscillations. At resonance with the standing wave, $\omega_0 = \omega_n$, the oscillations occur with a unit amplitude and frequency $\Omega_{\rm R} = 2g$. (It is the latter relation that identifies the parameter $g$ as the coupling strength between the qubit and the standing wave.)

At sufficiently large $t$, the previously neglected terms in $\Sigma(\omega)$ start to become important for the qubit dynamics [i.e., the second and third terms in the square brackets of Eq.~\eqref{eq:self-energy_result}]. 
The main effect of these terms is the decay of the Rabi oscillations.
To see the decay, note that---unlike $\omega_-$---the frequency $\omega_+$ is submerged into the continuum of phonon modes $\omega_n(\bm{k}_\perp)$, i.e., $\omega_+ > \omega_n$. The self-energy has an imaginary part in this frequency domain; the effect of the imaginary part amounts to the shift of the pole $\omega_+$ off the real axis\footnote{We expect such a treatment to correctly capture the dynamics on the qualitative level. Note, however, that at resonance, $\omega_0 = \omega_n$, ${\rm Im}\,\Sigma(\omega_+)$ is of the same order of magnitude as the spacing between two subsequent diffraction maxima $\sim \sqrt{\omega_{\rm diff}\,g}$. Because of this, a quantitatively correct description of the qubit dynamics would require a more accurate analysis of Eq.~\eqref{eq:solution_for_A}; this is beyond the scope of the present discussion.}, $\omega_+ \rightarrow \omega^\prime_+ - i\,{\rm Im}\,\Sigma(\omega_+)$.
The shift leads to the decay of the first term in  Eq.~\eqref{eq:ampl_oscill}, with a decay constant $\gamma_{\rm R} = {\rm Im}\,\Sigma(\omega_+)$.
Let us consider a qubit in resonance with the standing wave, $\omega_n = \omega_0$. In this case, $\omega_+ = \omega_n + g$ which---upon substitution into Eq.~\eqref{eq:self-energy_result}---results in
\begin{equation}\label{eq:rabi_decay}
    \gamma_{\rm R} \sim \sqrt{\omega_{\rm diff}\,g};
\end{equation}
a respective $t \sim 1 / \gamma_{\rm R}$ determines a time over which the Rabi oscillations cease. We note that the Rabi oscillations are well-resolved for $g \gg \omega_{\rm diff}$: a large number of oscillations, $\sim \sqrt{g / \omega_{\rm diff}} \gg 1$, occurs within the decay time $1/\gamma_{\rm R}$.

Curiously, the excited state population of the qubit does not fall to zero in our model (within the described above simplifying assumptions). Indeed, a second term in Eq.~\eqref{eq:ampl_oscill} remains non-vanishing even at $t \gg 1 / \gamma_{\rm R}$; at resonance,  we find $|A_e(t)|^2 = 1 / 4$ .  This peculiar behavior is a consequence of the closeness of the qubit frequency to the threshold of the phonon mode continuum $\omega_n$. The further relaxation happens either due to the qubit coupling to the off-resonant phonon modes with lower overtone numbers $n^\prime \leq n - 1$, or due to other mechanisms unrelated to the emission of phonons.

To conclude, we present estimates of parameters $\omega_{\rm diff}$, $g$, and $\gamma_{\rm R}$ for our experimental device. To estimate $\omega_{\rm diff}$ we use $a \simeq 300\,\mu$m, $\omega_n / 2\pi = 6$~GHz, and $v_\perp = 9.2$~km/s [to obtain the latter number, we use an equation for $v_\perp$ presented after Eq.~\eqref{eq:spectrum} and the material parameters presented in Sec.~\ref{sec:mat-params}]. Substituting these numbers in Eq.~\eqref{eq:omega_diff}, we obtain
\begin{equation}\label{eq:diff_broad_est}
    \omega_{\rm diff}/2\pi \simeq 20\,{\rm kHz}.
\end{equation}
Next, we use Eq.~\eqref{eq:g_standing_wave} to find $g$. To this end, we first estimate the electric field using the last entry of Table~\ref{tab:parameters}, $E_z \simeq 4.5\cdot10^{-2}$~V/m. As for the piezoelectric constant, we use the value $e_{33} = 0.52$~C/m$^2$, which we inferred in the main text from the phonon emission into a rough-backside chip [see Sec.~IIA]. Additionally, we use $b = 100$~$\mu$m for the chip thickness, $b_p = 1$~$\mu m$ for the thickness of the AlN layer, $n \simeq 110$ for the overtone number, and $\rho =3980$~kg/m$^3$ for the density of sapphire.  This leads to
\begin{equation}\label{eq:g_st_w_estimate}
    g / 2\pi \simeq 3\,{\rm MHz}. 
\end{equation}
Note that the coupling strength is small compared to the free spectral range $\nu_{\rm fsr} = v_l / 2b = 55\,$MHz. This justifies our approximation, in which we left a single resonant overtone $n$ in the expression for $\Sigma(\omega)$ [see the discussion before Eq.~\eqref{eq:self-energy}].
Finally, we find for the decay rate of the Rabi oscillations with the help of Eq.~\eqref{eq:rabi_decay}:
\begin{equation}\label{eq:gamma_R_estimate}
    \gamma_{\rm R} / 2\pi \simeq 240\,{\rm kHz}. 
\end{equation}
Expressions \eqref{eq:diff_broad_est}--\eqref{eq:gamma_R_estimate} need to be viewed as the order of magnitude estimates; thus, we conclude $\omega_{\rm diff} \sim 10$~kHz, $g/2\pi \sim 1$~MHz, and $\gamma_{\rm R}/2\pi \sim 100$~kHz.
\clearpage
\section{Splitting of the shear wave resonances}
The shear wave emission peaks in Fig.~2 of the main text are split by $\Delta\omega_{\rm split}\approx 2\pi \cdot 4\,{\rm MHz}$. In this section, we show that the splitting may originate from the misalignment between the device's $z$-axis and the $c$-axis of the sapphire crystal by $\Delta \theta \approx 0.15^{\circ}$, close to the axes alignment error in the specifications of the sample. 

\begin{figure}
    \centering
    \includegraphics[width=0.25\textwidth]{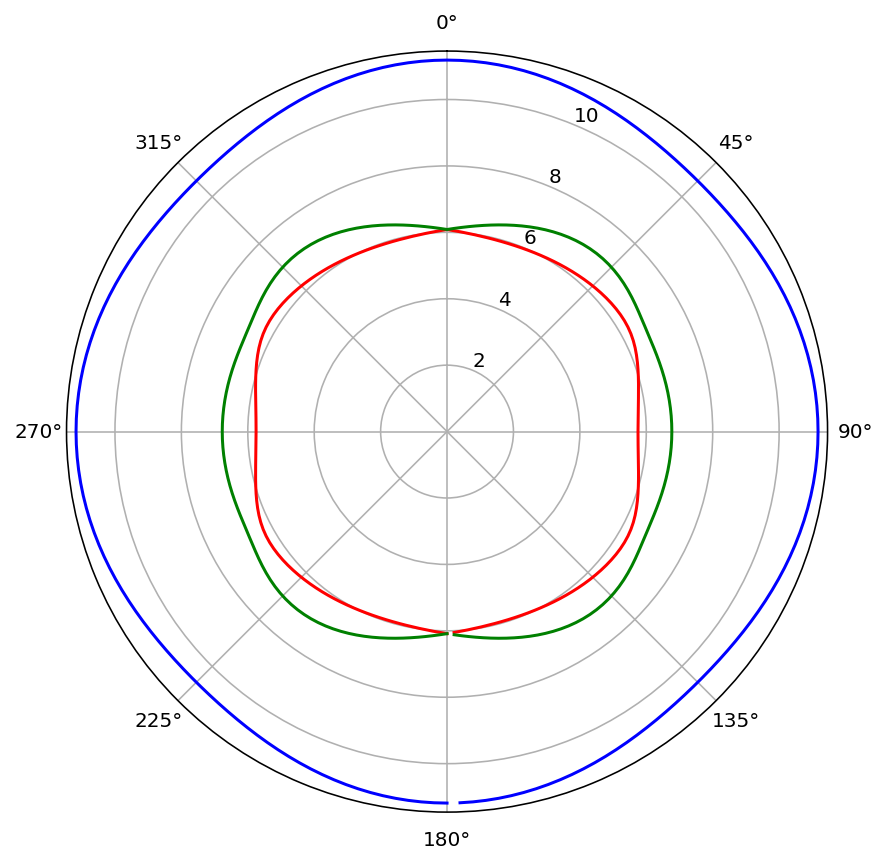}
    \includegraphics[width=0.75\textwidth]{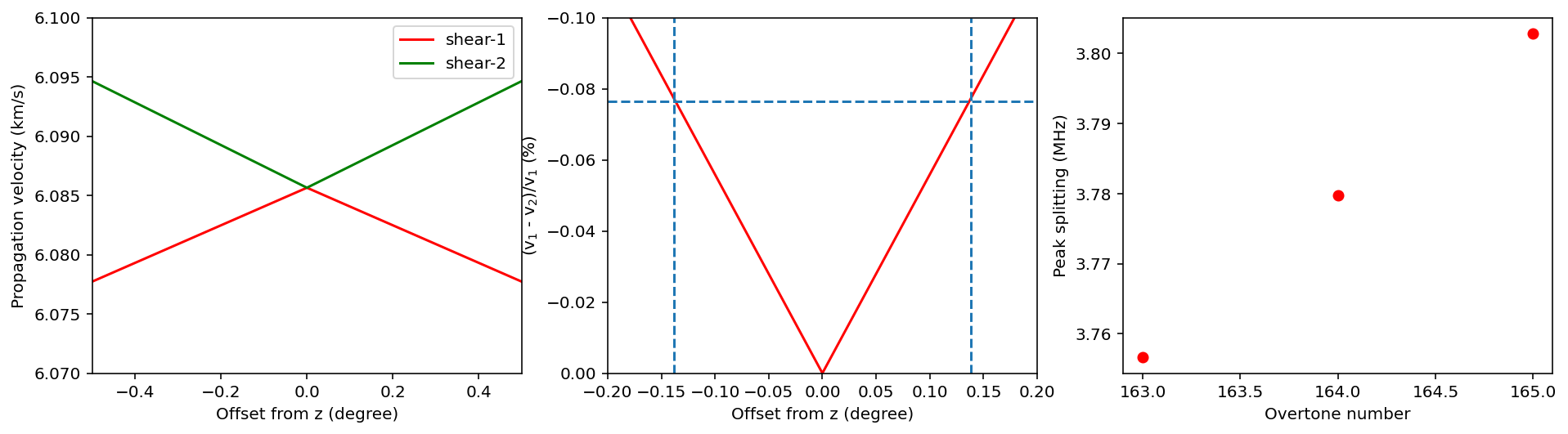}
    \caption{{\bf Velocity surfaces.} 
    A solution of Christoffel's equation in bulk sapphire along the Z-X crystal plane results in three velocities that depend on the direction of propagation  {($0^\circ$~corresponds to the propagation along the $c$-axis)}. In blue is the longitudinal velocity solution and in red and green are the two shear wave solutions in units of km/sec. 
    }
    \label{app:slowness-surface-aln}
\end{figure}

The waves are launched by the qubit in the $z$-direction which is perpendicular to the sapphire chip. If this direction coincided with the $c$-axis of the sapphire crystal, then the two shear waves would have the same velocities, and corresponding sets of the standing wave frequencies would be identical. The axes misalignment leads to a difference in velocities $\Delta v_{\rm sh}$ and thus the splitting of the standing wave frequencies. For the two shear waves close to the qubit frequency $\omega_0$ the splitting is given by
\begin{equation}\label{eq:split}
    \Delta \omega_{\rm split} = \omega_0 \frac{\Delta v_{\rm sh}}{v_{\rm sh}}.
\end{equation}
To find $\Delta v_{\rm sh}$, we start with the Christoffel wave equation for the displacement field $\vec{u}$ (the summation over the repeated indexes is assumed):
\begin{equation}\label{eq:christoffel}
    \rho\frac{\partial^2 {u}_{i}}{\partial t^2}-c_{ij,kl}\frac{\partial^{2}}{\partial r_{j}\partial r_{k}}u_{l}=0\quad\rightarrow\quad\rho\omega^{2}u_{i}-c_{ij,kl}k_{j}k_{k}u_{l}=0.
\end{equation}
Here $\rho$ is the density of sapphire and $c_{ij,kl}$ is the elastic stiffness tensor (see Sec.~\ref{sec:mat-params} for the valeues of $c_{ij,kl}$).
According to Eq.~\eqref{eq:christoffel}, we need to determine the eigenvalues of the matrix $M_{ij} = c_{ij,kl}n_jn_k / \rho$ to find $v^2$ for the three polarizations of the wave propagating in the direction $\vec{n}$.
The result of the numeric diagonalization of  $M_{ij}$ for arbitrary $\vec{n}$ is shown in Fig.~\ref{app:slowness-surface-aln}.
We do indeed see that the velocities of the two shear waves differ when $\vec{n}$ is not aligned with the $c$-axis. 

Let us now estimate the velocity splitting analytically. Were the sapphire $c$-axis perfectly aligned with the $z$-axis of the device, the matrix $M$ would be diagonal for $\vec{n} || \hat{z}$, $M_0 = {\rm diag}\{v^2_{\rm sh}, v^2_{\rm sh}, v^2_l \}$, where $v_{\rm sh} = \sqrt{{c_{44}}/ \rho}$ and $v_l = \sqrt{{c_{33}} / \rho}$ are the shear and longitudinal wave velocities, respectively (we use the standard notations $c_{44} \equiv c_{13,13}$ and $c_{33} \equiv c_{33,33}$). 
The misalignment of axes by $\Delta\theta \ll 1$ acts as a perturbation $\Delta M$ to the matrix $M_0$. To the first order in $\Delta \theta$, the perturbation is given by
\begin{equation}
\Delta M = \Delta\theta \Bigl[\frac{2c_{14}}{\rho}\left(\begin{array}{ccc}
-{\rm cos\,}\varphi & {\rm sin\,}\varphi & 0\\
{\rm sin\,}\varphi & {\rm cos\,}\varphi & 0\\
0 & 0 & 0
\end{array}\right)-\frac{c_{13}+c_{44}}{\rho}\left(\begin{array}{ccc}
0 & 0 & {\rm cos\,}\varphi\\
0 & 0 & {\rm sin\,}\varphi\\
{\rm cos\,}\varphi & {\rm sin\,}\varphi & 0
\end{array}\right)\Bigr],
\end{equation}
where $\varphi$ is the angle between the $x$-axis and the projection of the sapphire $c$-axis onto the $xy$-plane, $c_{14} \equiv c_{11,23}$, and $c_{13} \equiv c_{11,33}$. To find the splitting $\Delta v_{{\rm sh}}$, we apply the first-order degenerate perturbation theory to $M_0 + \Delta M$. It yields $\Delta v_{{\rm sh}} = \Delta\theta\,v_{{\rm sh}}\cdot 2|c_{14}|/c_{44}$ which, with the help of Eq.~\eqref{eq:split}, translates into
\begin{equation}
    \Delta\omega_{\rm split} = \omega_0\,\Delta\theta\,\frac{2|c_{14}|}{c_{44}} \simeq 2\pi\cdot 30\,{\rm MHz} \cdot \Delta \theta [^\circ].
\end{equation}
We used the values of the stiffness tensor components presented in Sec.~\ref{sec:mat-params} and $\omega_0 / 2\pi = 5.0\,{\rm GHz}$ for the qubit frequency. We see that the observed splitting, $\Delta\omega_{\rm split} \approx 2\pi \cdot 4\,{\rm MHz}$, is consistent with the axes misalignment by $\Delta \theta \approx 0.15^\circ$.

\clearpage
\section{Acoustic bound states}
In this section, we find the frequency spectrum of the acoustic bound states arising in a resonator with an appropriately deformed surface.
In our experiments, the role of the surface deformation is played by the shaped AlN transducer, see Fig.~\ref{fig:experimental-geometry}(e, f) (we shall dispense with the mismatch of the acoustic impedances of $\sim 20\%$ between the transducer and the acoustic resonator). 
First, we consider a transducer whose profile is smooth on the scale of the acoustic wavelength, see Sec.~\ref{sec:smooth}. Then, we find the bound states in the case of cylindrical transducer, see Sec.~\ref{sec:cylinder}.
Throughout the section, we focus on the bound states associated with the longitudinal waves to which the qubit couples the strongest. 

\subsection{Transducer with a smooth profile\label{sec:smooth}}
To describe the acoustic bound states associated with the longitudinal waves, we consider the Christoffel equation (Eq.~\eqref{eq:christoffel}) for the $z$-component of the displacement field $\vec{u}$:
    \begin{equation}\label{eq:christoffel_z}
        \partial_t^2{u}_{z} = v_{\perp}^{2} \nabla^2_\perp u_z + v_{l}^{2} \partial^2_z u_{z}.
    \end{equation}
Here we neglected $u_x$ and $u_y$ components of the displacement, which is justified for a smooth (on a scale of the acoustic wavelength $\lambda_{\rm ac}$) transducer. Velocities are given by $v_\perp^2 = [c_{44} + (c_{13} + c_{44})^2 / (c_{33} - c_{44})]/\rho$ and $v_l^2 = c_{33} / \rho$.
We can describe the shape of the transducer by specifying the dependence of the overall thickness of the chip on spatial coordinate, or $\thickness \equiv \thickness(\bm{r}_{\perp})$.
Provided the thickness varies adiabatically slowly, $|\partial \thickness /\partial \bm{r}_\perp| \ll 1$, it is possible to approximately separate variables in Eq.~\eqref{eq:christoffel_z} in the spirit of the Born-Oppenheimer approximation. To this end, we first solve Eq.~\eqref{eq:christoffel_z} for the wave motion in the $z$-direction at a given $\bm{r}_\perp$. This results in a set of standing waves $u_z \propto \cos{\pi n z / \thickness(\bm{r}_\perp)}$ indexed by the overtone number $n$ with frequencies $\omega_{z}(\bm{r}_\perp) = \pi n v_l / \thickness(\bm{r}_\perp)$. The slowness of the thickness variation guarantees that the overtone number $n$ is conserved. We can then use the ansatz
$u_{z}(\bm{r}_\perp,z) = \psi_n(\bm{r}_\perp) \cos{\pi n z/\thickness(\bm{r}_\perp)}$ in Eq.~\eqref{eq:christoffel_z}, which leads to the following Helmholtz equation for $\psi_n(\bm{r}_\perp)$:
    \begin{equation}
        \omega^{2} \psi_n(\bm{r}_\perp) = -v_{\rm \perp}^{2} \nabla^{2}_{\perp} \psi_n(\bm{r}_\perp) + \omega_{z}^{2}(\bm{r}_\perp) \psi_n(\bm{r}_\perp).
    \end{equation}
It is convenient to measure $\omega$ with respect to the standing wave frequency $\omega_n = \pi n v_l / \thickness$ of the HBAR in the region away from the transducer, where the device is nominally flat. To this end, we express $\omega^2 = \omega_n^2 + \varepsilon$, which leads to:
    \begin{equation}
        -v_{\rm \perp}^{2} \nabla_{\perp}^{2} \psi(\bm{r}_\perp) + (\omega_{z}^{2}(\bm{r}_\perp) - \omega_{n}^{2}) \psi(\bm{r}_\perp) = \varepsilon \psi(\bm{r}_\perp)
    \end{equation}
This equation is similar in form to the time-independent Schr\"{o}dinger equation. 
In this analogy, the topographic deformation can be thought of as a local potential $V(r_{\perp}) = (\pi n v_{l})^2\bigl[\frac{1}{\thickness^2(\bm{r}_\perp)} - \frac{1}{\thickness^2} \bigr]$. Since $\thickness(\bm{r}_\perp) \geq \thickness$, the potential is confining and thus gives rise to the bound states. 

To describe the structure of the bound states, we take $\thickness(\bm{r}_\perp) = \thickness + z(\bm{r}_\perp)$ and assume $z(\bm{r}_\perp) \ll \thickness$. Using the latter condition, we can expand $\frac{1}{\thickness(\bm{r}_\perp)^{2}} - \frac{1}{\thickness^{2}} \simeq - \frac{2}{\thickness^{3}} z(\bm{r}_\perp)$, which further simplifies the wave equation to
    \begin{align}
        -v_{\rm \perp}^{2} \nabla_{\perp}^{2} \psi_n(\bm{r}_\perp) - 2 \omega_{n}^{2} \frac{z(\bm{r}_\perp)}{\thickness} \psi_n(\bm{r}_\perp) = \varepsilon \psi_n(\bm{r}_\perp).  \label{eq:wave_final}
    \end{align}

So far, we have not made any assumptions about the shape of the transducer. Let us now apply Eq.~\eqref{eq:wave_final} to find the spectrum of the bound state frequencies for the dome-shaped transducer. The respective thickness profile is parabolic, $z(\bm{r}_\perp) = z_{0} (1-(r_{\perp}/r)^{2})$, where $z_{0}$ is the height of the dome and $r$ is the radius of its base. In terms of the dome's radius of curvature $\mathcal{R}=r^{2}/2z_{0}$, the wave equation has the form:
    \begin{equation}
        -v_{\rm \perp}^{2} \nabla_{\perp}^{2} \psi_n(r_{\perp}) + \frac{\omega_{n}^{2}}{\thickness \mathcal{R}} r_{\perp}^{2} \psi_n(r_{\perp}) = (\varepsilon + 2 \omega_{n}^{2} \frac{z_{0}}{\thickness}) \psi_n(r_{\perp})
    \end{equation}
It is similar to the Schr\"odinger equation for a harmonic oscillator. Using the solution of the harmonic oscillator problem, we find a set of transverse modes (indexed by the transverse wave number $m$) for each overtone number $n$  with frequencies:
    \begin{align}
        \omega_{n,m}^{2} &= \omega_{n}^{2} (1 - 2 z_{0}/\thickness) + \frac{2 \omega_{n} v_{\perp}}{\sqrt{\mathcal{R} \thickness }} (m + 1) = k_{n}^{2} v_{l}^{2} + k_{m}^{2} v_{\rm \perp}^{2}
    \end{align}
where $k_{n}^{2} = (\pi n/\thickness)^{2}(1-2z_{0}/\thickness)$ and $k_{m}^{2} = 2\omega_n v_\perp/\sqrt{\mathcal{R} \thickness}) (m + 1)$. Under the conditions $\lambda_{\rm ac} \ll \sqrt{{\cal R} \thickness}$ and $z_0 \ll \thickness$---which we assume to be well-satisfied---we can approximate $\omega_{n,m}$ by
\begin{equation}
    \omega_{n, m} = \omega_n (1 - z_0 / \thickness) + \frac{v_\perp}{\sqrt{{\cal R} \thickness}}(m + 1). 
\end{equation}
This shows that the spectrum of the bound state frequencies is equidistant, with the spacing $v_\perp / \sqrt{{\cal R}\thickness}$. We note, however, that the qubit couples to the modes with even $m$ only (the qubit's electric field has an approximate radial symmetry, so it does not couple to the modes with odd $m$ which are odd under the inversion). We thus find for the spacing between the consequent modes to which the qubit couples:
\begin{equation}
    \Delta \omega = \omega_{n,m+2} - \omega_{n,m} = \frac{2 v_\perp}{\sqrt{{\cal R} \thickness}}.
\end{equation}
Using the parameters $v_\perp = 9.2\,{\rm km/s}$
, ${\cal R} = 7.8\,{\rm mm}$, and $b = 100\,{\rm \mu m}$ corresponding to our device, we find $\Delta \omega = 2\pi \cdot 3.3\,{\rm MHz}$, close to the observed value $\Delta \omega = 2\pi\cdot (3.3-3.6)\,{\rm MHz}$.

The bound states are well-defined as long as $\omega_{n,m}$ lies below the threshold frequency $\omega_n$.
The latter marks the beginning of the continuous spectrum of waves (with an overtone number $n$) propagating in the flat part of the resonator. This leads to a bound for number $m$:
    \begin{equation}
        m + 1 < 2\pi \frac{z_0}{\thickness} \frac{v_l}{v_\perp} \frac{\sqrt{{\cal R} b}}{\lambda_{\rm ac}}.
    \end{equation}
From this condition, we conclude that {$15$ bound states} should be resolvable in the qubit spontaneous emission spectrum, close to the observed value.

\subsection{Cylindrical transducer\label{sec:cylinder}}
Next, we describe the acoustic bound states in the resonator with a cylindrical transducer.
Due to the sharpness the transducer's edge, the adiabatic approximation used in the previous section is no longer applicable. 
The quasi-discrete modes localized in the volume above the transducer are nonetheless long-lived. 
This is a consequence of the immense difference between the acoustic wavelength and the dimensions of the device: the diffraction losses are small as long as $r \gg \sqrt{\lambda_{\rm ac}\thickness}$, where $r$ is the radius of the transducer and $\thickness$ is the thickness of the resonator.

Under the condition\footnote{We focus on a practically relevant case $z_0 \sim \lambda_{\rm ac}$; for $z_0 \lesssim \lambda_{\rm ac}$ the condition reads $r \gg \sqrt{\lambda_{\rm ac} b \cdot (\lambda_{\rm ac} / z_0)}$.} $r \gg \sqrt{\lambda_{\rm ac}\thickness}$,
the bound states are almost entirely localized in the volume of the resonator above the transducer.
This means that we can effectively impose the zero displacement boundary condition at $r_\perp = r$ to find the bound state frequencies, i.e., $u_z(r_\perp = r) = 0$.
Focusing on the waves with an overtone number $n$ and keeping in mind the radial symmetry of the device, we look for the solution of the Christoffel equation of the form $u(\bm{r}_\perp, z) = R(r_\perp) \Phi(\varphi) \cos{\pi n z/(\thickness + z_0)}$, where $\Phi(\varphi)=e^{im\varphi}$ and  $R(r_\perp = r) = 0$. Substituting $u(\bm{r}_\perp, z)$ into Eq.~\eqref{eq:christoffel_z}, we find for $R(r_\perp)$:
    \begin{equation}\label{eq:R_eq}
        -\rho^{2} k_\perp^{2} R(r_\perp) = \left(r_\perp^{2} \partial_{r_\perp}^{2} + r_\perp \partial_{r_\perp} - m^{2} \right) R(r_\perp),
    \end{equation}
where $k_\perp^{2} = (\varepsilon + 2 \omega_{n}^{2} z_{0}/\thickness)/v_{\rm \perp}^{2}$. The solution of Eq.~\eqref{eq:R_eq}, which is regular at $r_\perp = 0$, is the Bessel function, $R_{m}(r_\perp) = J_{m}(k_\perp r_\perp)$. By imposing  $R(r_\perp = r) = 0$, we obtain the quantization condition for the transverse wavevector: $k_\perp^{m,l} = \mu_{m,l}/r$,  where $\mu_{m,l}$ is the $l$-th root of the Bessel function $J_{m}(x)$. The frequency of the mode $\{n,m,l\}$ is given by:
\begin{equation}
        \omega_{n,m,l} = \sqrt{\omega_{n}^{2} (1-2 z_{0}/\thickness) + [k_\perp^{m,l}]^2 v_{\rm \perp}^{2}} \approx \omega_n (1 - z_0 / \thickness) + \frac{\mu_{m,l}^2 v_{\rm \perp}^{2}}{2\omega_n r^2}.
\end{equation}
In contrast to the dome transducer, the modes in the cylinder are unequally spaced; the spacing increases with the radial number $l$. 

The bound states remain well-defined as long as $\omega_{n,m,l} < \omega_n$. This leads to the following condition for $l$ and $m$:
    \begin{equation}\label{eq:cond_bessels}
        \mu_{m, l}^{2} < \frac{8\pi^2 v_l^2}{v_{\rm \perp}^{2}} \frac{z_{0} r^2}{\thickness \lambda_{\rm ac}^2}.
    \end{equation}
We note that the qubit couples only to the modes with $m = 0$. Then, using Eq.~\eqref{eq:cond_bessels} and taking $r = 125\,{\rm \mu m}$, we find that $\approx 19$ bound states should be resolvable in the spontaneous emission spectrum for each overtone number $n$.

In Fig.~3 of the main text, we show the calculated frequency spectrum of the bound states by the vertical grey lines overlaid on the data.
We take $v_\perp = 9.2\,{\rm km/s}$, $v_l = 11.2\,{\rm km/s}$, and $\lambda_{\rm ac} = 2.2\,{\rm \mu m}$ corresponding to the qubit frequency $\omega_0 = 2\pi \cdot 5.0\,{\rm GHz}$. (Note that even a small uncertainty in $v_l$ of about $\sim 1\%$ may result in an overall shift of the bound state frequency spectrum by $\sim v_l/(2\thickness) \sim 50\,\rm{MHz}$ near $\omega_0$; in order to compare the theory and the data, we add a constant frequency offset which guarantees that the leftmost vertical line is aligned with the highest peak in the decay rate.)  Resonator thickness $\thickness = 100~\mu$m, radius $r = 125\,{\rm \mu m}$, and $z_0 = 1\,{\rm \mu m}$ are the same of the dome and the cylinder (the radius of curvature of the dome is ${\cal R} = r^2/2z_0 = 7.8\,{\rm mm}$).
We see that the cylinder has many more modes accessible at low $k_\perp$ at a given overtone number $n$, with uneven spacing. By comparison, the  
modes in the dome are evenly spaced; the set of the bound state frequencies is less dense than that for a cylinder. 

Finally, we find the coupling strength $g_{n,0,1}$ of the qubit to the principal mode in the cylinder. The normalized displacement field corresponding to the mode is given by
\begin{equation}
 u_z(\bm{r}_\perp, z) = \sqrt{\frac{2}{(\thickness + z_0)\,\pi r^2}}{\rm cos}\Bigl(\frac{\pi n z}{\thickness + z_0}\Bigr)\frac{J_0(\mu_{0,1} r_\perp / r)}{J_1(\mu_{1,0})}.
\end{equation}
With the help of Eq.~\eqref{eq:coupling_H}, we find for the coupling strength:
\begin{equation}
g_{n, 0, 1} = \frac{2e_{33}}{\sqrt{\hbar \rho \omega_{n, 0, 1} \thickness\,\pi r^2}} {\rm sin}^2\Bigl(\frac{\omega_{n, 0, 1} z_0}{2 v_l}\Bigr)\int d^2\bm{r}_\perp E_z(\bm{r}_\perp) \Bigl[\frac{J_0(\mu_{0,1} r_\perp / r)}{J_1(\mu_{1,0})}\Bigr]^2
\end{equation}
(we neglect $z_0 \ll \thickness$ in the normalization). Using the result of the HFSS simulation of the electric field (see Fig.~\ref{fig:app-e-fields-hfss}) and taking $e_{33} = 0.5\,{\rm C/m^2}$ (see the discussion in Sec.~IIA of the main text), we find $g_{n,0,1} \approx 2\pi \cdot 1\,{\rm MHz}$ for a coupling to the mode near the qubit frequency $\omega_0 = 2\pi \cdot 5.0\,{\rm GHz}$.
\clearpage

\section{Dynamics of the qubit coupled to acoustic modes in the dome}
\begin{figure}[h]
    \centering
    \includegraphics[width=0.80\textwidth]{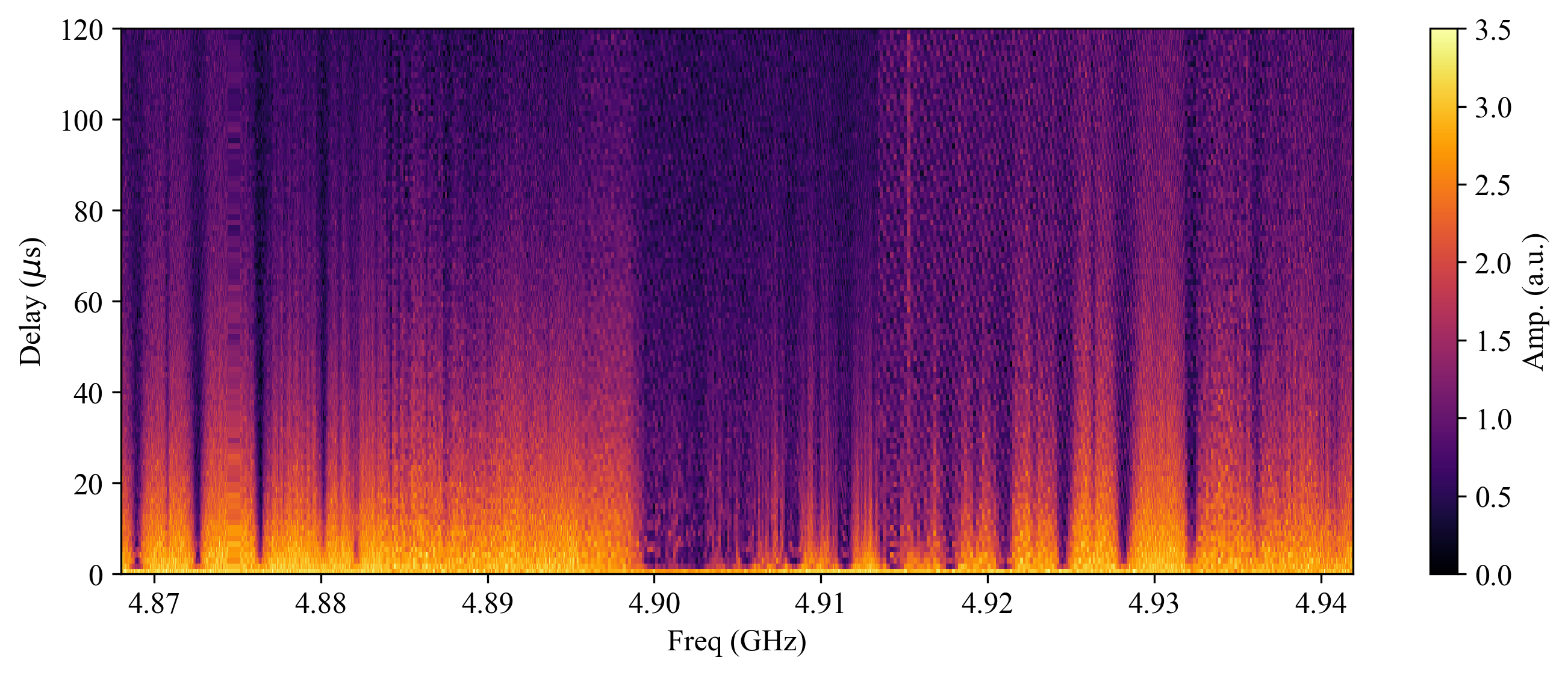}
    \caption{{\bf Dynamics of a qubit coupled to the dome-shaped resonator, as a function of the qubit frequency.} The vertical stripes correspond to a quick decay of an initially excited qubit; they occur at resonances between the qubit and the lossy modes localized in the dome  (i.e., the acoustic bound states).
    }
    \label{app:rabi-dome_1}
\end{figure}

\begin{figure}[h]
    \centering
    \includegraphics[width=0.5\textwidth]{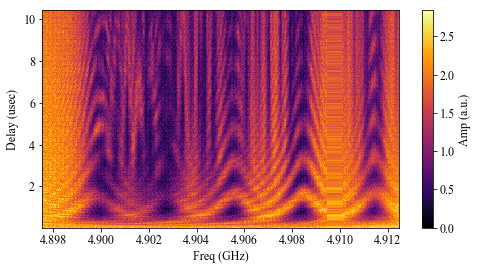}
    \caption{{\bf Close up of the early-time dynamics.} The excited state population of the qubit undergoes the vacuum Rabi oscillations at resonances with the acoustic bound states. For the shown resonances, the Rabi frequency is $\Omega_{\rm R} \approx 2\pi \cdot 0.6$~MHz.
    }
    \label{app:rabi-dome_2}
\end{figure}

Here, we present experimental results for the time-evolution of the qubit coupled to acoustic modes in the dome, see Figs.~\ref{app:rabi-dome_1} and \ref{app:rabi-dome_2}.
When the qubit frequency is close to a resonance with the frequency of an acoustic bound state, the qubit's excited state population undergoes vacuum Rabi oscillations, see Fig.~\ref{app:rabi-dome_2}.

\clearpage
\section{Material parameters}\label{sec:mat-params}
\subsection{Aluminum Nitride}
There are several relevant material parameters for the calculations described in our work. The piezoelectric strain tensor for aluminum nitride is
    \begin{equation}
    \overleftrightarrow{d} = \begin{pmatrix}
        0 & 0 & 0 & 0 & d_{15} & 0\\
        0 & 0 & 0 & d_{15} & 0 & 0\\
        d_{31} & d_{31} & d_{33} & 0 & 0 & 0
    \end{pmatrix}.
    \end{equation}
Here $d_{31} = -2.8$~pm/V, $d_{33} = 5.6$~pm/V (these values were found for the bulk AlN in Ref.~\cite{guy99}), and $d_{15} = 3.6$~pm/V~\cite{muensit99}.
The elastic stiffness matrix for AlN is
    \begin{equation}
    \overleftrightarrow{c}^{E} = \begin{pmatrix}
        c_{11} & c_{12} & c_{13} & 0      & 0      & 0\\
        c_{12} & c_{11} & c_{13} & 0      & 0      & 0\\
        c_{13} & c_{13} & c_{33} & 0      & 0      & 0\\
        0      & 0      & 0      & c_{44} & 0      & 0\\
        0      & 0      & 0      & 0      & c_{44} & 0\\
        0      & 0      & 0      & 0      & 0      & c_{66}
    \end{pmatrix},
    \end{equation}
where $c_{11} = 376$ GPa, $c_{12} = 129$ GPa, $c_{13} = 98$ GPa, $c_{33} = 353$ GPa, $c_{44} = 113$ GPa, and $c_{66}=\tfrac{1}{2}(c_{11}-c_{12})=124$ GPa \cite{deJong2015_elasticity}. 
The density of AlN is $\rho = 3255$~kg/m$^{3}$.
It can be convenient to combine the piezoelectric tensor $\overleftrightarrow{d}$ and the stiffness tensor to get the piezoelectric stress tensor $\overleftrightarrow{e} = \overleftrightarrow{d} \cdot \overleftrightarrow{c}$:
    \begin{equation}
    \overleftrightarrow{e} = \begin{pmatrix}
        0 & 0 & 0 & 0 & e_{15} & 0\\
        0 & 0 & 0 & e_{15} & 0 & 0\\
        e_{31} & e_{31} & e_{33} & 0 & 0 & 0
    \end{pmatrix},
    \end{equation}
where $e_{31}=d_{31}(c_{11}+c_{12})+d_{33}c_{13}=-0.86~\rm{C/m}^{2}$, $e_{33}=2d_{31}c_{13}+d_{33}c_{33}=1.43~\rm{C/m}^{2}$, and $e_{15}=d_{15}c_{44}=0.41~\rm{C/m}^{2}$.
We note that the values of the piezoelectric constants for thin films depend on the growth method and conditions, and vary across different experiments \cite{muralt1999, guy99, iqbal18}. For instance, Ref.~\cite{muralt1999} reports $e_{31} = -1.0\,\rm C/m^2$. The values of $e_{ij}$ listed above can only be viewed as the estimates with the accuracy of $\sim 10\%$. In Table \ref{tab:parameters}, we round $e_{33}$ and $e_{15}$ to the first decimal place.

The piezoelectric effect stiffens the elasticity tensor, such that 
    \begin{equation}
        \overleftrightarrow{c}^{D} = \overleftrightarrow{c}^{E} + \overleftrightarrow{e}^{T} \cdot \bigl(\varepsilon_0 \overleftrightarrow{\varepsilon}\bigr)^{-1} \cdot \overleftrightarrow{e}.
    \end{equation}
Here $\varepsilon_0$ is the vacuum permittivity and $\overleftrightarrow{\varepsilon}$ is the dielectric permittivity tensor. The latter is given by 
    \begin{equation}
    \overleftrightarrow{\varepsilon} = \begin{pmatrix}
        \varepsilon_{11} & 0 & 0\\
        0 & \varepsilon_{11} & 0\\
        0 & 0 & \varepsilon_{33}
    \end{pmatrix}.
    \end{equation}
For AlN, the values are $\varepsilon_{11}=8.2$ and $\varepsilon_{33}=9.7$. 

Combining these, we get the effective stiffness tensor for AlN is
    \begin{equation}
    \overleftrightarrow{c}^{D} = \begin{pmatrix}
        379.1 & 132.1 & 89.7 & 0      & 0      & 0\\
        132.1 & 379.1 & 89.7 & 0      & 0      & 0\\
        89.7 & 89.7 & 374.9 & 0      & 0      & 0\\
        0      & 0      & 0      & 115.3 & 0      & 0\\
        0      & 0      & 0      & 0      & 115.3 & 0\\
        0      & 0      & 0      & 0      & 0      & 123.5
    \end{pmatrix}.
    \end{equation}
The longitudinal and shear wave velocities for the propagation in the $z$-directions are $v_l^{(\rm AlN)} = \sqrt{c^D_{33}/\rho} = 10.7\,\rm km / s$ and $v_{\rm sh}^{\rm (AlN)} = \sqrt{c^D_{44}/\rho} = 5.9\,\rm km /s$, respectively. 

\subsection{Sapphire}
The elastic stiffness tensor for sapphire is~\cite{wachtman60}
    \begin{equation}
    \overleftrightarrow{c} = \begin{pmatrix}
        c_{11} & c_{12} & c_{13} & c_{14} & 0      & 0\\
        c_{12} & c_{11} & c_{13} & -c_{14}& 0      & 0\\
        c_{13} & c_{13} & c_{33} & 0      & 0      & 0\\
        c_{14} & -c_{14}& 0      & c_{44} & 0      & 0\\
        0      & 0      & 0      & 0      & c_{44} & c_{14}\\
        0      & 0      & 0      & 0      & c_{14}& c_{66}
    \end{pmatrix},
    \end{equation}
where $c_{11} = 496.8$ GPa, $c_{12} = 163.6$ GPa, $c_{13} = 110.9$ GPa, $c_{14} = -23.5$ GPa, $c_{33} = 498.1$ GPa, $c_{44} = 147.4$ GPa, and $c_{66}=\tfrac{1}{2}(c_{11}-c_{12})=166.6$ GPa. Its density is $\rho = 3980$ kg/m$^{3}$. The longitudinal and shear wave velocities for the wave propagating in the $z$-direction are
$v_l = \sqrt{c_{33}/\rho} = 11.2\,\rm km / s$ and $v_{\rm sh} = \sqrt{c_{44}/\rho} = 6.1\,\rm km /s$, respectively. Note that these velocities are close (within $5\%$) to the respective velocities for AlN. We neglect the difference between the velocities in the two media in all of our estimates.

\bibliography{references}